\documentclass[letterpaper]{raa}            

\usepackage{graphicx,times,amssymb}             
\input{epsf.sty}                        
\input{psfig.sty}                       
\usepackage[totalwidth=480pt,totalheight=680pt]{geometry}

\usepackage{enumerate}

\def\cm{\,{\rm cm}}

\def\ergscm2 {erg\,s$^{-1}$cm$^{-2}$}

\def\cm2 {cm$^{-2}$}

\def\aap {A\&A}
\def\apj {ApJ}
\def\apjs {ApJS}
\def\prd {Phys. Rev. D}
\def\prc {Phys. Rev. C}
\def\mnras {MNRAS}

\def\apss{ApSS}
\def\nat{Nature}
\def\sovast{Soviet Ast.}
\def\araa{ARA\&A}

\begin{document}

   \title{Quark-Novae in  Neutron Star-White-Dwarf  Binaries:\\
 A  model for   luminous (spin-down powered)  sub-Chandrasekhar-mass Type Ia Supernovae ?}

  \volnopage{Vol.0 (200x) No.0, 000--000}      
   \setcounter{page}{1}          

   \author{Rachid Ouyed
      \inst{1}
   \and Jan Staff
      \inst{2}
   }

  \institute{Department of Physics and Astronomy, University of Calgary, 
2500 University Drive NW, Calgary, Alberta, T2N 1N4 Canada; {\it rouyed@ucalgary.ca}\\
        \and
             Department of Physics and Astronomy, Louisiana State University,
202 Nicholson Hall, Tower Dr., Baton Rouge, LA 70803-4001, USA\\
   }

   \date{Received~~2012 month day; accepted~~2012~~month day}

\abstract{ We show that appealing to a  Quark-Nova (QN) in a tight  binary system   containing a massive
neutron star  and a CO white dwarf (WD), a Type Ia explosion could occur.  The  QN ejecta  collides
with the WD driving a shock that   triggers  Carbon burning under degenerate conditions (the QN-Ia). 
The conditions in the compressed low-mass  WD ($M_{\rm WD} < 0.9M_{\odot}$) in our model mimics those of
 a Chandrasekhar mass WD. The spin-down luminosity from the QN compact remnant (the quark star) provides
     additional power that makes the QN-Ia light-curve brighter  and broader
    than a standard  SN-Ia with similar $^{56}$Ni yield.  
    In QNe-Ia, photometry and spectroscopy are not necessarily linked since the kinetic energy of the
    ejecta has a contribution from  spin-down power and nuclear decay. 
     Although QNe-Ia  may not obey the Phillips relationship, their brightness and their      relatively ``normal looking"   light-curves
      means they could  be included in the cosmological sample. Light-curve fitters 
    would be confused by the discrepancy between spectroscopy at peak and photometry   and would correct for it
      by effectively brightening or dimming the QNe-Ia apparent magnitudes. Thus
    over- or under-estimating the true magnitude of these spin-down powered SNe-Ia. 
       Contamination of QNe-Ia  in samples of SNe-Ia used for cosmological analyses could   systematically bias measurements of cosmological parameters  if QNe-Ia are numerous enough at high-redshift.   The strong mixing induced by  spin-down wind combined with the low $^{56}$Ni yields in QNe-Ia   means that these  would  lack a secondary maximum in the $i$-band despite their luminous nature.   We discuss possible QNe-Ia progenitors.
}

   \authorrunning{R. Ouyed  \& J. Staff }            
   \titlerunning{Quark-Novae in NS-WD Binaries: spin-down powered Type Ia Supernovae ?}  

   \maketitle

\section{Introduction}
\label{sec:introduction}

Despite their astrophysical significance, as a major contributor to cosmic nucleosynthesis and as distance indicators in observational cosmology, Type Ia supernovae (SNe-Ia)  lack theoretical understanding.  The evolution leading to explosion and its  mechanisms are among  the unknowns.
The  consensus is that  type Ia supernovae  result from thermonuclear explosions of carbon-oxygen (CO) white dwarfs (WDs; Hoyle \& Fowler 1960; Arnett 1982).  The explosion proper is generally thought to be triggered when the WD approaches (for accretion) or exceeds (for a merger) the Chandrasekhar mass, and the density and temperature become high enough to start runaway carbon fusion.   Detonation  models have been proposed for C-burning in the WD interior (Arnett 1969; Nomoto 1982) as well as deflagration models (Woosley \& Weaver 1986).
 A delayed detonation transition (Khokhlov 1991a) may be needed to better replicate observations.

The nature of the progenitors of SNe-Ia is debated. 
Explosion models of SNe-Ia currently discussed in the literature include explosions of Chandrasekhar mass WDs
and its variants (Khokhlov  1991b; Gamezo et al. 2005; Livne et al. 2005; R\"opke \& Niemeyer 2007; Jackson et al. 2010; Plewa 2007; Jordan et al. 2008; Meakin et al. 2009;  Bravo et al. 2009 to cite only a few), 
  explosion of  super-massive WDs (e.g. Pfannes et al. 2010 and references therein), and  of sub-Chandrasekhar WDs (Woosley et al. 1980; Nomoto 1982; Livne \& Glasner 1991; Livne \& Arnett 1995; Fink et al. 2010).

In the single degenerate (SD) scenario,  if mass transfer is too slow, novae occur, which appear to remove as much mass as was accreted (Townsley \& Bildsten 2004). If it is faster, H burns stably, but only a small range of accretion rate  avoids expansion and mass-loss (Nomoto et al. 2007).  The lack of H in spectra of SNe-Ia is often seen as troublesome for SD progenitor models.
On the other hand, in the double-degenerate (DD) scenario,  mergers of WDs   could give rise to SNe-Ia (Webbink 1984; Iben\& Tutukov 1984) and could  naturally explain the lack of H.   Both SD and DD scenarios may allow super-Chandrasekhar SNe-Ia. If the WD is spun up by accretion to very fast differential rotation (with mean angular velocity of order a few radians per second on average), then the WD may exceed the physical Chandrasekhar mass by up to some tenths of a solar mass before reaching explosive conditions in the central region (Yoon \& Langer 2005).
Merger simulations did not result in an explosion (e.g. Saio\&Nomoto 2004) rather they indicate that 
an off-centre ignition causes the C and O to be converted to O, Ne, and Mg, generating a gravitational collapse rather than a thermonuclear disruption
 (Nomoto \& Iben 1985). This is the so-called accretion-induced
collapse (AIC) to a NS where  C is not ignited explosively but quietly, yielding a faint explosion and a NS remnant instead of a SN-Ia (see also Stritzinger et al. 2005).

\subsection{sub-Chandrasekhar mass models}
\label{sec:sub-Chandra}

Theoretical and numerical (hydrodynamical) studies have previously shown that sub-Chandrasekhar mass WDs with an overlying helium shell (accreted from
a companion) can undergo a double-detonation which could lead to a SN-Ia (Woosley et al. 1980; Nomoto 1982; Glasner \& Livne 1990; Livne \& Glasner 1991; Livne \& Arnett 1995; Fink et al. 2007; Fink et al. 2010).   In these models a layer of accreted helium ($\sim$  0.1-0.2$M_{\odot}$) is built either by burning accreted hydrogen to helium or by accretion of helium from a helium-rich donor (Woosley\&Weaver 1986;  Woosley\&Weaver 1994; Ivanova\&Taam 2004). When the pressure at the base of the helium layer reaches a critical threshold, it detonates driving a shock into the core of the WD. This causes a second detonation, resulting in a flame propagating outward from the core (or near it), destroying the WD. 
   In edge-lit models, the mass of the WD must increase during the pre-supernova evolution to $\sim 0.9$-$1.1 M_{\odot}$ to explain typical SN-Ia luminosities (e.g. Woosley\&Kasen 2011).  This strong constraint on the WD mass is due to the fact that core densities $> 2.5\times 10^7$  g cm$^{-3}$ are required  for the detonation to produce enough radioactive Nickel (Sim et al. 2010) and to survive Nova-like
outbursts at the high accretion rate which actually shrink the WD mass. Specifically, the WD mass should be  at least 0.9 $M_{\odot}$ at the time of the SN-Ia (to produce an amount of $^{56}$Ni within the range of normal SNe).

 Although physically realistic, the double-detonation sub-Chandrasekhar model may suffer from the fact that
even with a very low mass helium layer ($\sim 0.05M_{\odot}$) their spectroscopic signatures are
 not characteristic of observed SNe-Ia (Kromer et al. 2010; see also Ruiter et al. 2011). 
However, it has recently been argued that the model might be capable of producing a better match to observations, depending on details regarding the manner in which the accreted helium burns (e.g. Fink et al. 2010). 
 It has also been suggested that a more
 complex composition of the helium layer may lead to a better agreement with observations but
 this remains to be  confirmed.    More recent 1-dimensional simulations show that only the hottest (i.e.,  with initial luminosity of $\sim L_{\odot}$), most massive WDs considered with the smallest helium layers, show reasonable agreement with the light-curves and spectra of common Type Ia supernovae (Woosley \&Kasen  2011).

In the DD scenario,  the less massive WD may be disrupted
into a  disk from which the more massive WD
accretes at a constant rate near the gravitational Eddington limit. 
Others find that the less massive WD is transformed into a hot, slowly rotating, and radially extended envelope supported by thermal pressure (e.g. Shen et al. 2012 and references therein). It was  found that the long-term evolution of the merger remnant is similar to that seen in previous calculations; i.e. an off-center burning eventually yielding a high-mass O/Ne WD or a collapse to a NS, rather than a Type Ia supernova (see also Dan et al. 2011).
 On the other hand, van Kerkwijk et al. (2010) consider the viscous evolution of mergers of equal mass WDs in which both WDs are tidally disrupted (see also Yoon et al. 2007; Lor\'en-Aguilar et al. 2009; Pakomar et al. 2011). 
 The resulting remnant has a temperature profile that peaks at the center (and is fully mixed), unlike remnants in which only one WD is disrupted, which have a temperature peak in material at the edge of the degenerate core. 
 
  The sub-Chandrasekhar mass WD mergers (van Kerkwijk et al. 2010) lead
   to a cold remnant ($\sim 6\times 10^8$ K) with central densities $\sim 2.5\times 10^6$ gm cm$^{-3}$. However, 
   accretion of the thick ``disk"  leads to compressional heating 
   resulting in an increase in the central temperature to $\sim 10^9$ K and densities $\sim 1.6\times 10^7$ gm cm$^{-3}$.
   These conditions they argue could  ignite the remnant centrally  with the nuclear runaway inevitable. 
 In this scenario, van Kerkwijk et al. (2010) argued that SNe-Ia result from mergers of CO WDs, even those with sub-Chandrasekhar total mass.    
 Badenes\&Maoz (2012) find a remarkable agreement between the total WD merger rate and the SN-Ia rate;  but  not enough close binary WD systems to reproduce the observed Type Ia SN rate in the classic DD scenario.
Apart from the consistency between SN-Ia rates and total WD merger rates, sub-Chandrasekhar explosions may have the advantage of producing the correct chemical stratification (Sim et al. 2010), without resorting to the  delayed detonation mechanism (Khokhlov 1991) needed by super-Chandrasekhar models. We note that  these simulations begin with the binary components close enough that strong mass transfer immediately sets in once the calculation is begun. In contrast, Dan et al. (2011) emphasize the importance of beginning such simulations at larger orbital separations and instead find tidal disruption at a much larger radius with correspondingly less violence.

We would like to mention other alternative progenitor scenarios to produce Type Ia supernova explosions, which are not restricted to the ignition of a CO WD near the Chandrasekhar mass.    One scenario involves  tidal disruption of white dwarfs by moderately massive black holes (Rosswog et al. 2009a) and 
 another involves  a shock-triggered thermonuclear explosion  from the collision of two WDs (Rosswog et al. 2009b). See also Milgrom\&Usov (2000)
  for Type Ia  explosions triggered by gamma-ray bursts.
For a detailed discussion on the open issue of SN-Ia progenitors, we refer interested reader to several reviews (e.g., Branch et al. 1995; Renzini 1996; Livio 2000). Overall, the models described above, and those listed in the reviews above, differ in their assumptions about initial conditions, ignition processes, whether the explosion involves subsonic deflagration or not, and other details, and they have a varying success in explaining basic observations of SNe-Ia.
  A common feature of the models is that all of them involve, in one way or another, the detonation mode of burning.  However, the lack
 of convincing solutions to the progenitor(s) of SNe-Ia leaves room for alternative. Here we present a new channel for Type Ias (SNe-Ia)
    by appealing    to a Quark-Nova explosion (hereafter QN; Ouyed et al. 2002; Ker\"anen et al. 2005) in a close NS-WD (CO) binary system.
     Under appropriate  conditions, C-burning is triggered by shock compression and heating from the relativistic QN ejecta  (QNE) 
 impacting the WD leading to a Type Ia explosion. Hereafter, we refer to these QN-triggered type Ias  as QNe-Ia.

  The basic picture  of the QN is that a  massive NS   converts 
explosively to a quark star (Ouyed et al. 2002; Ker\"anen et al. 2005).
Such an explosion can happen if the NS reaches the quark deconfinement density via spin-down or accretion (Staff et al. 2006) and subsequently undergoes a phase transition to the conjectured more stable strange quark matter phase (Itoh 1970; Bodmer 1971; Witten 1984; see also Terazawa 1979), resulting in a conversion front that propagates toward the surface in the detonative regime
(Niebergal et al. 2010) -- {\it a  hypothesis} we adopt  in this paper
(as in previous work) based on preliminary 1D simulations. The outcome is ejection of  the NS's outermost layers at  relativistic speeds.   The outer layers  are ejected  from an expanding thermal fireball (Vogt et al. 2004;
Ouyed et al. 2005)  which  allows for ejecta with kinetic energy, $E_{\rm QN}^{\rm KE}$, in the  $10^{52}$ erg range.
In previous papers, we introduced the QN as a model for superluminous SNe (Leahy\&Ouyed 2008; Ouyed\&Leahy 2012), discussed 
 their photometric/spectroscopic  signatures (Ouyed et al. 2012) as well as their nuclear/spallation signatures from the interaction  of the ultra-relativistic neutrons with the preceding SN shells and surroundings (Ouyed et al. 2011c; see also Ouyed 2012).
 We also explored conditions for QNe to  occur in binaries with
 applications to gamma-ray bursts (GRBs)  (Ouyed et al. 2011a\&b).

 Here we present a model in the context of QNe occurring in a NS-WD system
  and show how  luminous sub-Chandrasekhar mass Ia explosions  could in principle occur.
  This paper bears similarities to those of Ouyed et al. 2011b,
 but considers more carefully both the interaction between the QNE
 and the WD  and considers the implication of the spin-down luminosity of the QN compact
 remnant  (the quark star)  on the resulting light-curve. In Ouyed et al. 2011b 
 we explored both  the relativistic and non-relativistic degenerate regime while here we focus solely on
  the relativistic regime and consider only $M_{\rm WD} > \sim 0.5M_{\odot}$. 
       The main difference between QNe-Ia  and standard SNe-Ia are: (i) A QN-Ia involves the detonation
  of a sub-Chandrasekhar mass WD ($M_{\rm WD} < 0.9M_{\odot}$) in a close NS-WD (CO) binary system with orbital
  separation $< 10^{10}$ cm.  This hints at specific progenitors as discussed in this paper;
  (ii)  Burning  in QNe-Ia  occurs 
  following impact by the relativistic QNE. The compression and heating of the WD mimics burning conditions (densities and temperature)
   reminiscent of those in Chandrasekhar mass models although the CO WD in our model is truly in a sub-Chandrasekhar mass regime;
    (iii) In addition to $^{56}$Ni decay,  spin-down power from the QN compact remnant (the quark star) provides an additional energy
    source that powers the explosion. This additional energy source is unique to our model and has the potential of  altering the shape (i.e. morphology)
     of the light-curve.
  Readers who wish to understand the essential differences and/or differentiating 
predictions of our model compared to standard SNe-Ia are referred to \S \ref{sec:trigger}.

    The paper is organized as follows: In \S 2, we give a brief description of the QN. 
     In \S 3 we describe the  collision between the QNE and the WD
      and explore conditions for C-detonation to occur in NS-WD systems experiencing a QN. In this section,
       we explain how the QN can lead to a thermonuclear runaway in the companion WD. Here,
       we discuss the resulting nuclear products. 
        The spectrum  and the 
      light-curve are  discussed  in   \S \ref{sec:lightcurve&spectrum}. In particular,   we investigate how spin-down luminosity
      alters the resulting light-curve and discuss plausible QNe-Ia candidates among peculiar SNe-Ia.  In \S 5
 we present possible QNe-Ia progenitors and their occurrence rates.  In \S 6 we discuss plausible  QN-Ia connection
 to massive star formation and its implications to cosmology and Dark Energy.
Specific predictions and a conclusion are given in \S 7. 

\section{The Quark-Nova}
 
  \subsection{The exploding neutron star}
  \label{sec:Mc}

As in Staff et al. (2006), we assume deconfinement density of $\rho_{\rm c} = 5\rho_{\rm N}\simeq 1.25\times 10^{15}$ g cm$^{-3}$
where $\rho_{\rm N}=2.85\times 10^{14}$ g cm$^{-3}$ is the nuclear saturation density.
  For the APR equation of state (Akmal, Pandharipande, \& Ravenhall 1998), which we adopt in this paper, a  static configuration (i.e. non-rotating NS) of $M_{\rm NS,c.}\sim 1.8M_{\odot}$ reaches $\rho=\rho_{\rm c}$ in its core, thus prone  to the QN explosion.  Stiffer EoSs  (e.g. Ouyed\&Butler 1999) extend the critical NS mass to higher values ($M_{\rm NS,c.}\sim 2M_{\odot}$) while
  softer EOS (e.g. BBB2; Baldo, Bombaci, \& Burgio 1997) give lower values ($M_{\rm NS,c.}\sim 1.6M_{\odot}$). We note that all of these EoS
   allow for NSs with maximum masses higher than the $M_{\rm NS,c.}$. The QN effectively reduces
    the maximum mass allowed by a given EoS to $M_{\rm NS,c}$.  Naturally, rapidly rotating
    configurations will increase the mass limit.~\\

There are two possible paths to reaching deconfinement in the core of a NS:\\

(i) Via spin-down if the NS is born with a mass  above $M_{\rm NS,c}$ but fully recycled ($< 2$ ms);
the fast rotation decreases the core density below the $\rho_{\rm c}= 5\rho_{\rm N}$ limit.  
 Staff et al. (2006) considered the parameter-space  in mass, magnetic
field and  spin-period to  identify how long  such a NS would take to
reach the quark deconfinement  density. They found that NSs  with mass  $> M_{\rm NS,c}$,
spin period $\sim 2$ ms and magnetic field $\sim 10^9$G ($10^8$ G)  will  reach deconfinement 
density in $\tau_{\rm QN} < 10^8$ ($10^{10}$)
years  due  to  the  spin-down  effect  from  dipole radiation, leading
to an increase in the star's central density to the quark deconfinement limit. 
 
(ii) If the NS is born massive enough (very close to the critical NS mass $M_{\rm NS,c.}$) and
  mildly or slowly rotating. In this case, a slight increase in its
mass by accretion  will push its core
density above the deconfinement value which triggers the explosive instability.
 The combined effect of increase in the core density   from the added
 mass and its decrease from NS spin-up lead  to an overall increase
  of a few percents of the NS core density.  This scenario requires
  the NS mass  (core density) to be within a few percents or $M_{\rm NS,c}$ (or $\rho_{\rm c}$)
   to explode as a QN when it accretes.

  The mass limit, $M_{\rm NS,c}$, set by the QN 
  means that  NSs heavier than $M_{\rm NS,c}$ should not exist (if QNe were to occur) in nature 
  in contradiction  with the recent observations of a $2M_{\odot}$ NS (Demorest et al. 2010).  
  However, $M_{\rm NS,c}$ could be made to exceed $\sim 2M_{\odot}$ if we set the deconfinement density
  above $5\rho_{\rm N}$ as is the case  for the APR EoS  whose maximum gravitational mass exceeds $2M_{\odot}$.  
  It is also possible that  NSs more massive than $M_{\rm NS,c}$ are really quark or hybrid stars
  (that is, the quark matter EoS should be sufficiently stiff to support such a mass)\footnote{While it is almost certain that the (u,d,s) phase, if it exists inside NSs, cannot be a free gas of quarks (\"Ozel et al. 2010; Weissenborn et al. 2011), an interacting phase of quarks still appears to be consistent with the recent finding of a $2M_{\odot}$ NS (Demorest et al. 2010).}. 
  Heavy quark stars may exist, so long as the quark superconducting gap and strong coupling corrections are taken into account (Alford et al. 2007).
  The BBB2 EOS is too soft and provides a maximum gravitational mass of $\sim 1.9M_{\odot}$ but 
  if  the observed massive NS is really a quark or hybrid star, we cannot rule out BBB2 in this way, and its inclusion is still useful.

\subsection{The QN compact remnant: the Quark star}

In the QN model, we assume that hot quark matter in the Color-Flavor-Locked (CFL) phase is the true ground state of matter at high density (Alford et al. 1999). This is a superconducting phase that is energetically favored at extremely high densities and low temperatures. In this phase u, d, and s quarks pair, forming a quark condensate (a superfluid) that is antisymmetric in color and flavor indices. This state is reached by the QN compact remnant (a
Quark star in the CFL phase) as it cools below a few tens of MeV.   The initial QS surface magnetic field is  of the order of  $10^{14}$-$10^{15}$ G (Iwazaki 2005);  we adopt $10^{14}$ as a fiducial value. 

Spin-down of the QN compact remnant due to magnetic braking 
 can naturally lead to the launching of a secondary outflow in the form of a pair wind.
 The corresponding spin-down (lower case subscript $sd$) luminosity is
$L_{\rm sd}\sim 6.4\times 10^{43}\ {\rm erg\ s}^{-1}\ B_{\rm QS, 14}^2 P_{\rm QS, 10}^{-4} (1+ t/\tau_{\rm sd})^{-5/3}$
      where $\tau_{\rm sd}\simeq 100\ {\rm days} \ B_{\rm QS, 14}^{-2} P_{\rm QS, 10}^{2}$ (Staff et al. 2008; see Contopoulos \& Spitkovsky, 2006
       for spin-down power for an aligned rotator). Here
       the QS magnetic field is given in units of $10^{14}$ G and the period in units of 10 ms. I.e. a rotational energy (an additional energy source)
       of $\sim 2\times 10^{50} P_{\rm QS, 10}^{-2}$         
        not present in any of the standard (i.e. $^{56}$Ni powered) models of Type Ia SNe; the QS moment of inertia is set to $2\times 10^{45}$ g cm$^2$. 
        The implications to QNe-Ia light curves (with plausible deviations from the Phillips relationship) are presented in  \S \ref{sec:lightcurve}.

\subsection{The Quark Nova ejecta (QNE)}

The QN proper (i.e. the explosion) will  happen  on timescales of milliseconds (Ouyed et al. 2005; Niebergal et al. 2010)
ejecting the outermost layers of the parent NS (Ker\"anen et al. 2005) at relativistic speeds with an average
Lorentz factor $\Gamma_{\rm QN}\sim 10$.
     The evolution of the QNE from point of explosion is  given in appendix B in Ouyed\&Leahy (2009) with the 
    QNE density $\rho_{\rm QNE}$ at a distance $r=a$ from the point of
explosion derived from a combination of mass conservation and thermal spreading of the QN 
ejecta thickness, $\Delta r$. This gives 
\begin{equation}
\label{eq:rhoQNE}
\rho_{\rm QNE} \sim 1.8\times 10^6\ {\rm  g\ cm}^{-3}\times \frac{\rho_{0,14} \Delta r_{0, 4} }{a_{9}^{9/4} M_{\rm QN, -3}^{1/4}}\ ,
\end{equation}
where  $\rho_{\rm 0}$, the QNE density at explosion radius,  is given in units 
of  $10^{14}$ g cm$^{-3}$; for a typical ejecta mass $M_{\rm QNE} \sim 10^{-3} M_{\odot}$ ejecta.  
  The  thickness at ejection is $\Delta r_{0}\sim 10^4$ cm or $\Delta r_{0, 4}\sim 1$
  in units of $10^4$ cm; the distance from the explosion point (later
  to be defined as the binary separation), $a_{9}$, is given in units of $10^{9}$ cm.
  
\subsection{Specific observational signatures of QNe}

While this paper explores observational signatures of  QNe going off in  binary systems, we briefly 
  mention some specific and unique signatures of QNe going off in isolation. 
In particular, dual-shock QNe (i.e. QNe going off a fews days to a few weeks following the preceding type II SN explosion)
offer the most promising observables. The interaction of the QNE with the preceding SN ejecta
 leads to the re-energization of the SN shell which should manifest itself  in the optical as a ``double-hump" 
 lightcurve with the first, smaller, hump corresponding to the core-collapse SN proper and the second hump to the re-energized SN ejecta (Ouyed et al. 2009).  A strong contender for the double-humped lightcurve is  the  super-luminous supernova  SN2006oz as reported in Ouyed\&Leahy (2012; see also Ouyed et al. 2012).  
 
 Besides the re-energization of the preceding  ejecta from the Type II explosion, the extremely neutron-rich relativistic QNE
  leads to spallation of the innermost layers of the SN shell thus destroying 
   $^{56}$Ni while forming sub-Ni elements. One distinguishable feature
  of this interaction is the production of $^{44}$Ti at the expense of $^{56}$Ni (see Ouyed et al. 2011c)
   which results in Ni-poor (i.e sub-luminous), Ti-rich type II SNe and the proposal of   Cas-A as
   a plausible dual-shock QN candidate (Ouyed et al. 2011c); the QN imprint in Cas A might have been
    observed (Hwang \& Laming 2012).         The neutron-rich QN ejecta as it expands away from the NS
was shown to make mostly $A>130$ elements (Jaikumar et al. 2007).    
              We can thus combine photometric and spectroscopic signals that are specific to the QN -- and should thus be model-independent -- to come up with a plausible, observable candidate. Spectroscopically, the dual-shock QN will exhibit strong $\gamma$-ray signatures from $^{44}$ Ti and  from $A>130$ elements.  Combining the spectroscopic signal, with the photometric ``double hump" of dual-shock QN light curve, gives a very specific signature of how a QN will appear to observers. 
              In addition, the gravitational wave (GW) signal from the preceding SN  and the subsequent QN should be discernible in case of asymmetric explosions (Staff, Jaikumar, \& Ouyed 2012). GW observatories currently in planning may be able to detect this predicted dual-GW signals and may offer first glimpses of QNe in the near future.

   \section{QN-triggered Carbon Burning}
  \label{sec:trigger}

   Now let us consider a binary system with a massive   NS and a CO WD (hereafter MNS-COWD) where the NS experiences
   a QN episode.   We ask at 
  what distance  from the NS should the WD be located  when the QN
  goes off in order to  ignite  Carbon under  degenerate conditions?
 For this purpose, we adopt a WD mass of $0.6M_{\odot}$ representative of the empirical mean of
mass of WDs (e.g. Tremblay\&Bergeron 2009;  Kepler et al. 2006) with a mean density of $\sim 5\times 10^6$ g cm$^{-3}$ (e.g. Even\&Tohline 2009)
 -- we only consider the relativistic degenerate regime in this paper  which puts a lower limit on the
 WD mass we consider in this work.
 The WD  mass-radius relationship for this  regime is (Padmanabhan 2001)
  \begin{eqnarray}
 \frac{R_{\rm WD}}{R_{\odot} } &\simeq& \frac{0.011}{\mu_{\rm WD,2}} \left(\frac{M}{M_{\rm Ch}}\right)^{-1/3}\times f(M_{\rm WD})^{1/2}\ ,
 \end{eqnarray}
 where $f(M_{\rm WD}) = (1 - (\frac{M_{\rm WD}}{M_{\rm Ch}})^{4/3})$
 and $M_{\rm Ch.} = 1.435 M_{\odot}/\mu_{\rm WD, 2}^{2}$  the Chandrasekhar limit with $\mu_{\rm WD}$  the mean molecular weight in units of 2.
 
  \subsection{The QNE-WD collision}
  \label{sec:collision}

 When the QNE encounters the  WD (i.e. a number density jump $n_{\rm QNE}/n_{\rm WD}$),
 a reverse shock (RS) is driven into the cold QNE, while a forward shock (FS) propagates into
 the cold higher density WD material (see Appendix).   Therefore, there are four regions separated
by the two shocks in this system: (1) unshocked cold WD matter, (2) forward-shocked hot WD matter, (3) reverse-shocked QNE,
and (4) unshocked cold QNE.  From  the   shock jump conditions (see Appendix) one can 
  show that   the $n_{\rm QNE}/n_{\rm WD}  \simeq \Gamma_{\rm QNE}^2$ condition separates the
   Newtonian RS regime ($n_{\rm QNE}/n_{\rm WD}  > \Gamma_{\rm QNE}^2$) from the relativistic
   RS regime ($n_{\rm QNE}/n_{\rm WD}  < \Gamma_{\rm QNE}^2$).  
     When the RS is 
  Newtonian it  converts only a very small fraction of the kinetic energy into thermal energy; in this case the Lorentz
  factor of region 2 (the shocked WD material) relative to the rest frame of the WD (i.e. region 1; also an external observer) is  $\Gamma_2\simeq \Gamma_{\rm QNE}$.   The relativistic RS  limit 
  is the regime where most of the kinetic energy of the QNE is converted to thermal energy by the shocks (in this
  case $\Gamma_2\simeq (n_{\rm QNE}/n_{\rm WD})^{1/4}\Gamma_{\rm QNE}^{1/2}/\sqrt{2}$). 
For a recent analytical  formulation of relativistic shocks we
refer the reader to  Uhm (2011; and references therein).

  A relativistic RS is the critical condition to substantially heat the QNE. But 
 what matters in our case  is  the FS, which compresses and heats the WD. 
 After RS crossing of the QNE, the FS starts to decelerate, so
 $\Gamma_2$ is a decreasing function with time and radius.  More importantly, if by the time  the FS 
  has decelerated (i.e. reaches the non-relativistic regime) it has travelled deep enough inside the WD
   then this raises  the possibility of detonating the WD with the FS.  
    The time it takes the RS to cross the QNE (shell) is $\tau_{\rm cros}\simeq \Delta R_{\rm QNE} \Gamma_{\rm QNE} (n_{\rm QNE}/n_{\rm WD})^{1/2}/c$ (e.g. Sari\&Piran 1995); $c$ is the speed of light. To a first order, since 
   the QNE has a thickness of $\sim$ 100-1000 km at a distance 
   of $\sim 10^9$-$10^{10}$  cm from the explosion (see appendix B  in Ouyed\&Leahy 2009) and taking $n_{\rm QNE}/n_{\rm WD}\sim \Gamma_{\rm QNE}$,
    we get $\tau_{\rm cros}\sim 0.17\ {\rm s}\times \Delta R_{\rm QNE, 500} \Gamma_{\rm QNE, 10}^2$; $\Delta R_{\rm QNE, 500}$
    is in units of 500 km.  Thus, the RS lasts for a short period of time.  During this time the FS would have reached
     depths of at least of the order of $\Delta R_{\rm QNE}$ (since $\Gamma_{\rm FS}\simeq 2 \Gamma_{\rm RS}$). I.e. 
    depths of the order of a few thousands kilometers could be reached by the FS under appropriate conditions.
    We can define a corresponding critical   WD density  as (combining eq.(\ref{eq:rhoQNE}) and $n_{\rm QNE}/n_{\rm WD}  \simeq \Gamma_{\rm QNE}^2$)
   \begin{equation}
    \rho_{\rm WD, c} \sim 2\times 10^{5}\ {\rm g\ cm}^{-3}\  \times \frac{\rho_{0,14} \Delta r_{0, 4}}{\Gamma_{\rm QNE, 10}^2 a_{9}^{9/4} M_{\rm QN, -3}^{1/4}} \  ,
   \end{equation}
   where the QNE Lorentz factor is in units of 10. 
   To convert from number density to mass density, we take $\mu_{\rm WD}\sim 14$ and $\mu_{\rm QNE}\sim 1$ as the mean molecular
   weight for the WD and QNE, respectively (the QNE remains dominated by neutrons even after the end of the r-process; Jaikumar et al. 2007).  Equation above 
    is relevant if the radius at which the density reaches the critical value is smaller than the deceleration radius (which is the RS crossing radius). In general, and to a first approximation, we arrive at similar results by assuming that the non-relativistic stage would be reached by the FS   when the initial  energy of the QNE equals roughly the rest mass energy of the WD being shocked. 
  Our treatment of the  propagation of the FS shock is very simplified and whether
it can propagate that deep in the WD 
 remains to be confirmed by detailed numerical simulations of the QNE-WD interaction.
    To carry on with our investigation, we assume that some of our QNE the RS becomes
    non-relativistic deep inside the WD.  Under the right conditions 
    heating and compression could lead to Carbon ignition close to the core (i.e. at $R_{\rm WD}< R_{\rm WD,c}$).

  We note that the WD is always substantially heated regardless of whether the RS
is relativistic or Newtonian.
 The energy gained by the WD is  an important  portion of the QN kinetic energy, $E_{\rm WD, th.}\sim E_{\rm QN}^{\rm KE}\times \Omega_{\rm WD}$, where $\Omega_{\rm WD}= R_{\rm WD}^2/(4 a^2)$ is the solid angle subtended by the WD. Or,
\begin{equation}
E_{\rm WD, th.}
 \sim 4.7\times 10^{48} \ {\rm erg}~ \frac{\zeta_{\rm QN} E_{\rm QN, 52}^{\rm KE}}{a_{10}^2} \frac{f(M_{\rm WD})/0.68}{\mu_{\rm WD, 2}^{10/3}  M_{\rm WD,0.6}^{2/3}}\ ,
\end{equation}
 where we  made use  of the generalized  mass-radius relation for  WDs described earlier with $f(0.6M_{\odot})\sim 0.68$.
  The factor $\zeta_{\rm QN} < 1$ is relevant to cases where the thermal energy gained by the WD from heating by the QN shock is less than 100\%;  since lower values of  $\zeta_{\rm QN}$ are easily compensated by higher $E_{\rm QN}^{\rm KE}$
   values, hereafter we take $\zeta_{\rm QN} = 1$.
If (see \S \ref{sec:compression}  below) compression and thermalization of  the WD occurs 
 before ignition and burning takes place efficiently, then 
 the average temperature per nucleon of the shocked and thermalized WD is 
\begin{equation}
\label{eq:Tswd}
T_{\rm SWD} \sim 9.5\times 10^9 \ {\rm K} ~\frac{E_{\rm QN, 52}^{\rm KE}}{a_{9}^2}
\frac{f(M_{\rm WD})/0.68}{  \mu_{\rm WD, 2}^{7/3}  M_{\rm WD, 0.6}^{5/3}}\ .
\end{equation}

    \subsection{Shock  compression and carbon ignition/burning}
    \label{sec:compression}

 For $\Gamma_{\rm QNE} >>1$ the  density in the shocked WD material  is (see Appendix)
\begin{equation}
\label{eq:densityjump}
 \frac{\rho_{\rm SWD}^{\prime}}{\rho_{\rm WD}}\simeq \Gamma_2\times  \left( 4\Gamma_2 + 3\right)  \ ,
 \end{equation}
 or   $\rho_{\rm SWD}^{\prime}/\rho_{\rm WD}\propto \Gamma_2^2$
 where $\rho_{\rm SWD}^{\prime}$ is the density of the shocked WD material  in the  WD (i.e. observer's) frame. 
  We note that even for 
 a non-relativistic RS,  $\Gamma_2$ can be as high, or even higher than  $\Gamma_{\rm QNE}$ (e.g. Zhang \& Kobayashi 2005).
 
 In principle, the WD compression ratio can reach values of tens or  hundreds. 
  For our fiducial values,   the above translates to  (an order of magnitude estimates of the compression ratio of) 
 \begin{equation}
 \frac{\rho_{\rm SWD}^{\prime}}{\rho_{\rm WD}}\sim 
\left\{
 \begin{array}{rl}
430 &\mbox{if $ \rho_{\rm WD}  \le \rho_{\rm WD, c}$\ {\rm since}\ $\Gamma_2=\Gamma_{\rm QNE}$ }
\\
< 217  &\mbox{if $ \rho_{\rm WD}  > \rho_{\rm WD,c}$\ {\rm since}$\ \Gamma_2< 2.3  \Gamma_{\rm QNE} $}
\end{array}
\right.
\end{equation}   
In general then it is not unrealistic to assume that solutions can be found where the shocked WD might be compressed to  average 
   densities of $\sim 10^8$ g cm$^{-3}$ in its core and average densities of $\sim  10^7$ g cm$^{-3}$ in its surface layers. 
 While the highest compression will most likely be achieved  in the WD surface layers,   these will most likely experience minimal heating. 
   Thus, ignition and burning (if they occur successfully) we speculate will most likely be triggered in regions deeper than $R_{\rm WD,c}$ where higher temperatures are reached.    
   
       Correctly modelling the process that leads to successful ignition in our model requires 
    a more elaborate treatment (i.e. detailed numerical simulations) of the shock
    including necessary physics such as neutrino losses,  diffusive processes,  
    turbulence and so forth (e.g. Dursi \& Timmes 2006).    Here we can only  provide  very qualitative 
    arguments that allow us to speculate  that a suitable shock (i.e. with a relativistic FS travelling
    deep inside the WD) could in principle ignite the WD successfully: 
   The nuclear burning time scale  (Woosley et al. 2004) behind the shock   in our case can be
  estimated to be $\tau_{\rm nuc}\sim 1.153\times 10^{-20} \ {\rm s}\ (10^{10}\ {\rm K}/T_{\rm SWD})^{22} (10^8\ {\rm g\ cm}^{-3}/\rho_{\rm SWD})^{3.3}$.
  The time scale on which the WD can react, its dynamical time scale, is $\tau_{\rm dyn.} = (G \rho_{\rm WD})^{-1/2}\sim 1.7\ {\rm s}\ (5\times 10^6\ {\rm g\ cm}^{-3}/\rho_{\rm WD})^{1/2}$.
  Only in scenarios where  $\tau_{\rm dyn.}$ is much shorter than the burning time scale $\tau_{\rm nuc}$  would the WD respond fast enough to quench burning by a reduction of density  and temperature.  Appreciable burning  will   take place if  $\tau_{\rm nuc}< \tau_{\rm dyn.}$  or if $T_{\rm SWD} >  10^9\ {\rm K}\times (\rho_{\rm SWD}/10^8\ {\rm g\ cm}^{-3})^{-2.8/22}$.   Combined with eq.(\ref{eq:Tswd}) this gives
  \begin{equation}
  \label{eq:anuc}
  a_{9} <   a_{\rm nuc,9}\sim 3.1 \times {(E_{\rm QN, 52}^{\rm KE})}^{1/2}\frac{\left(f(M_{\rm WD})/0.68\right)^{1/2} }{  \mu_{\rm WD, 2}^{7/6}  M_{\rm WD,0.6}^{5/6}\rho_{\rm SWD,8}^{-2.8/44}}\ ,
  \end{equation}
  where the shocked WD density $\rho_{\rm SWD,8}$ is in units of $10^8\ {\rm g\ cm}^{-3}$.
  Note the very weak dependence of  $a_{\rm nuc, 9}$ on $\rho_{\rm SWD}$.
    To continue with our investigation we assume that the condition $\tau_{\rm cros.} < \tau_{\rm nuc.} < \tau_{\rm dyn.}$ is met. 
    I.e., the ignition and burning timescale are to a first order shorter than  dynamical timescale.

 \begin{figure*}
\begin{center}
\begin{tabular}{ccc}
\includegraphics[scale=0.43]{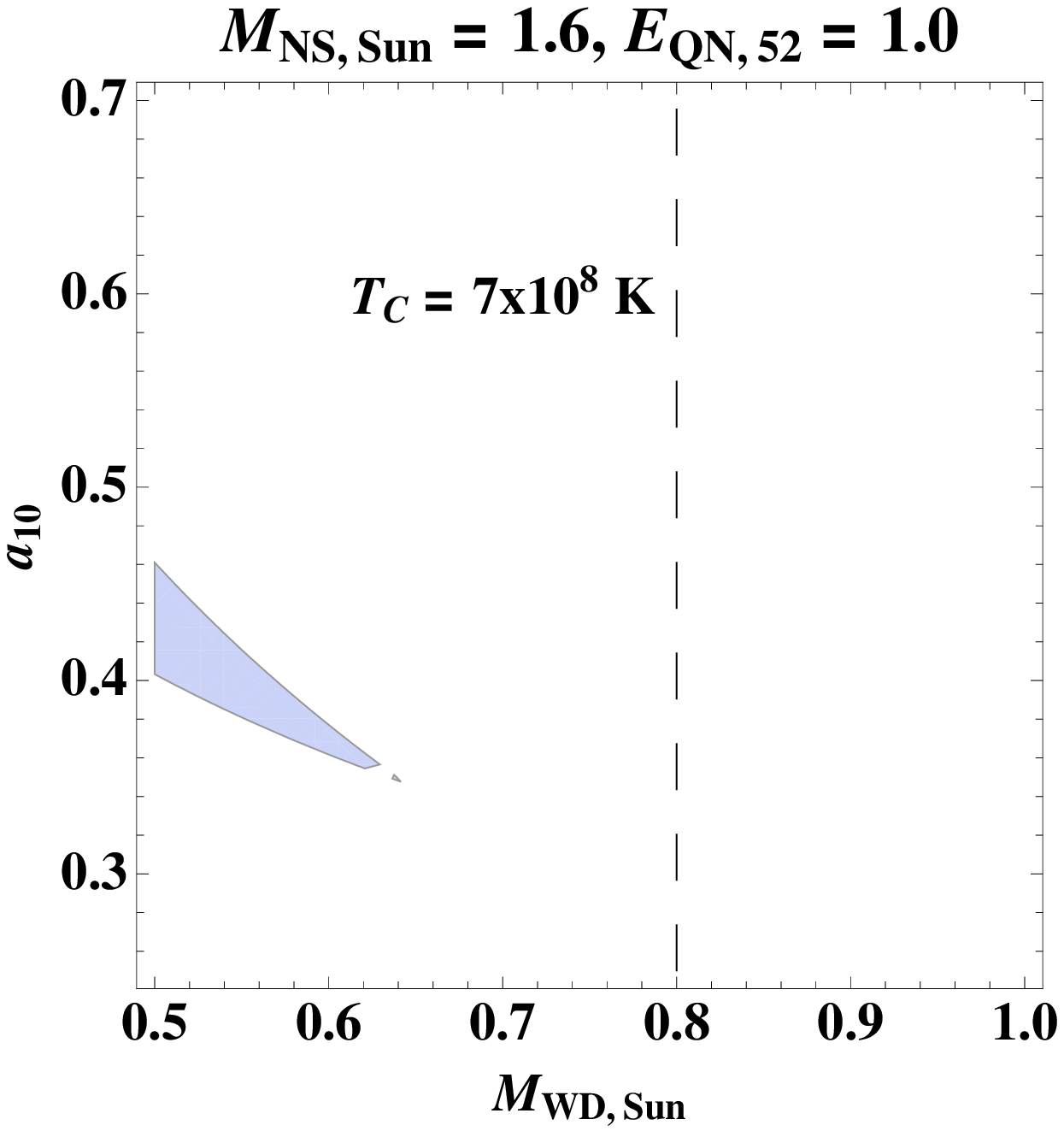} & \includegraphics[scale=0.43]{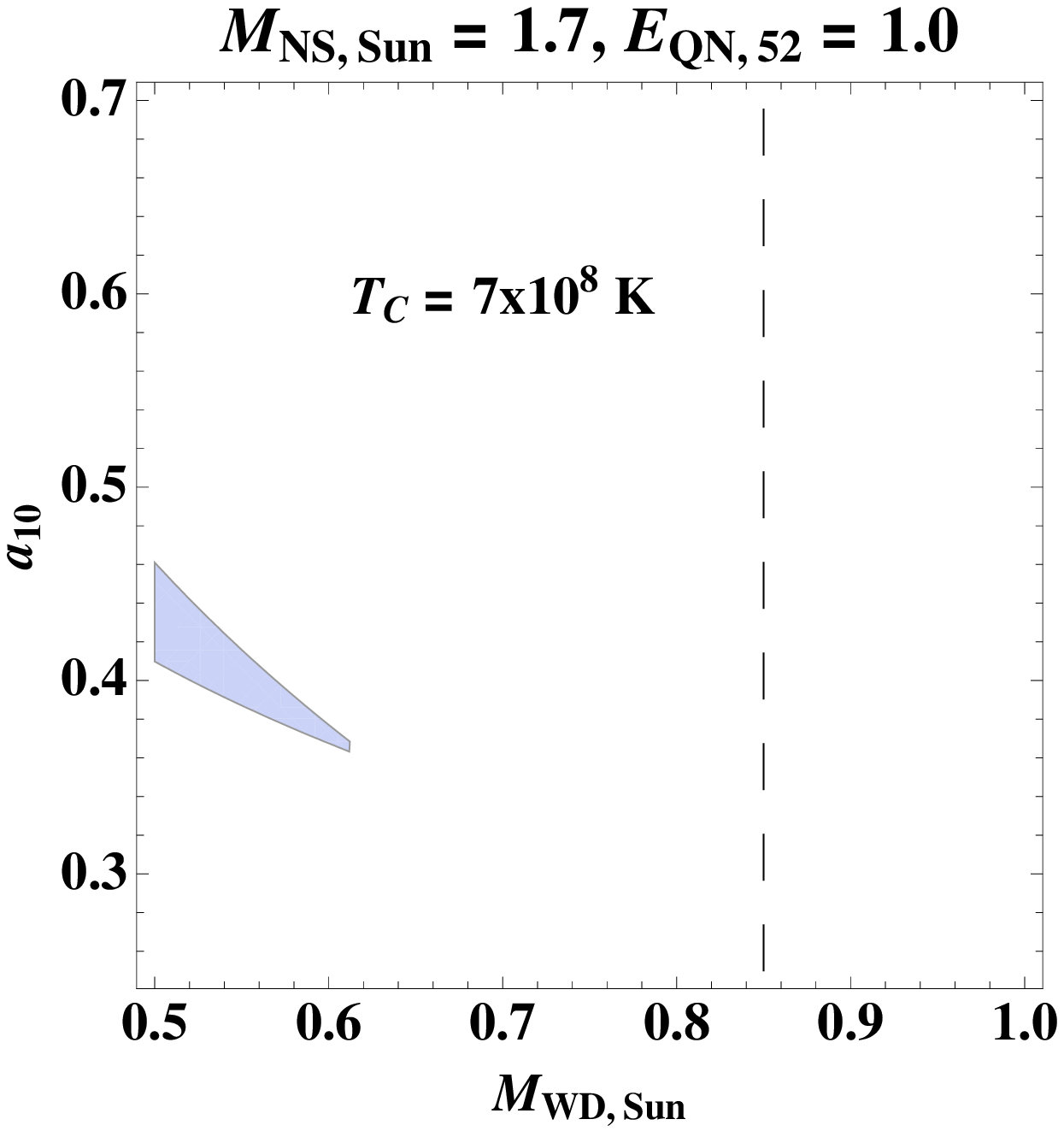}]& \includegraphics[scale=0.45]{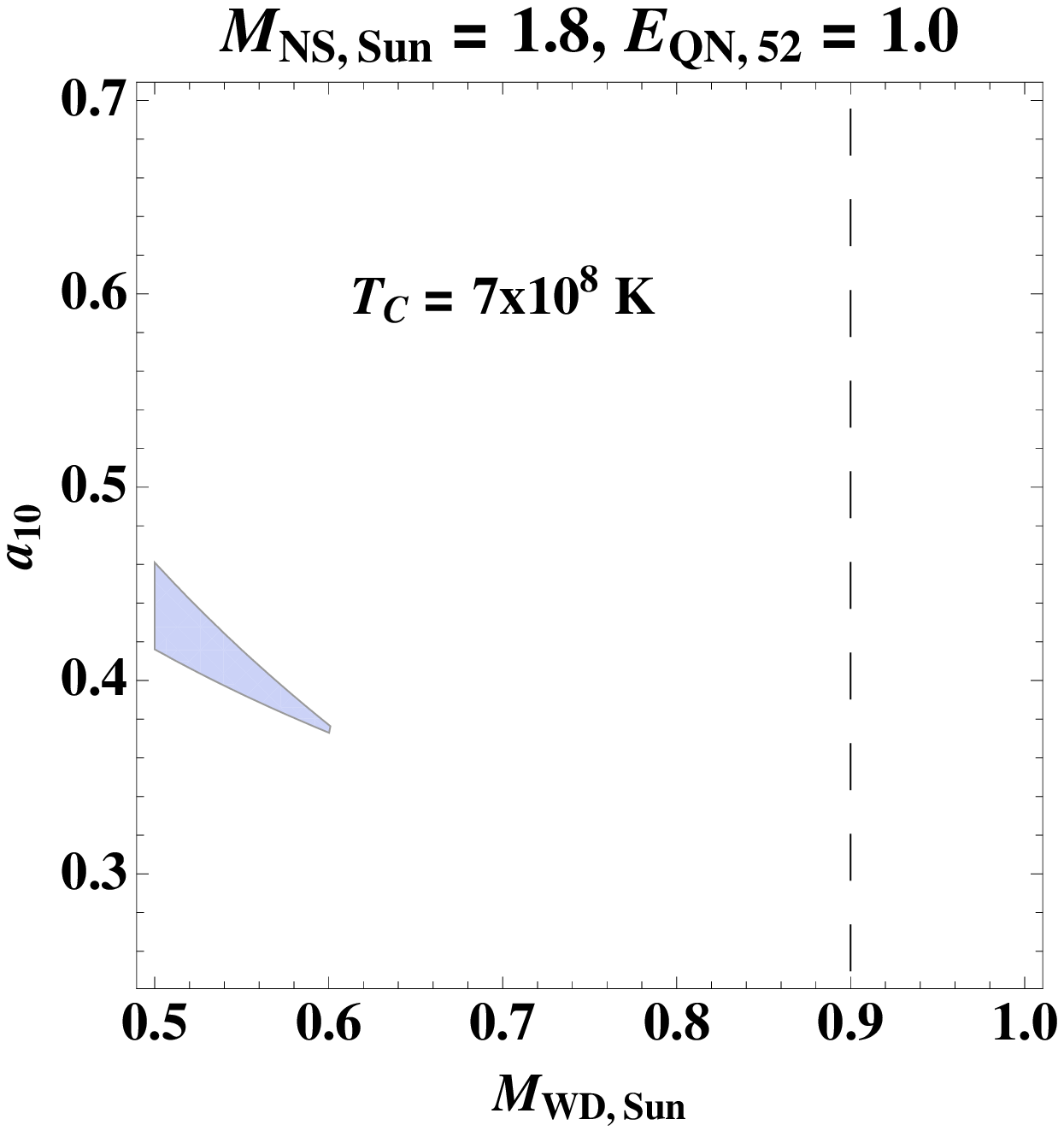} \\
\includegraphics[scale=0.43]{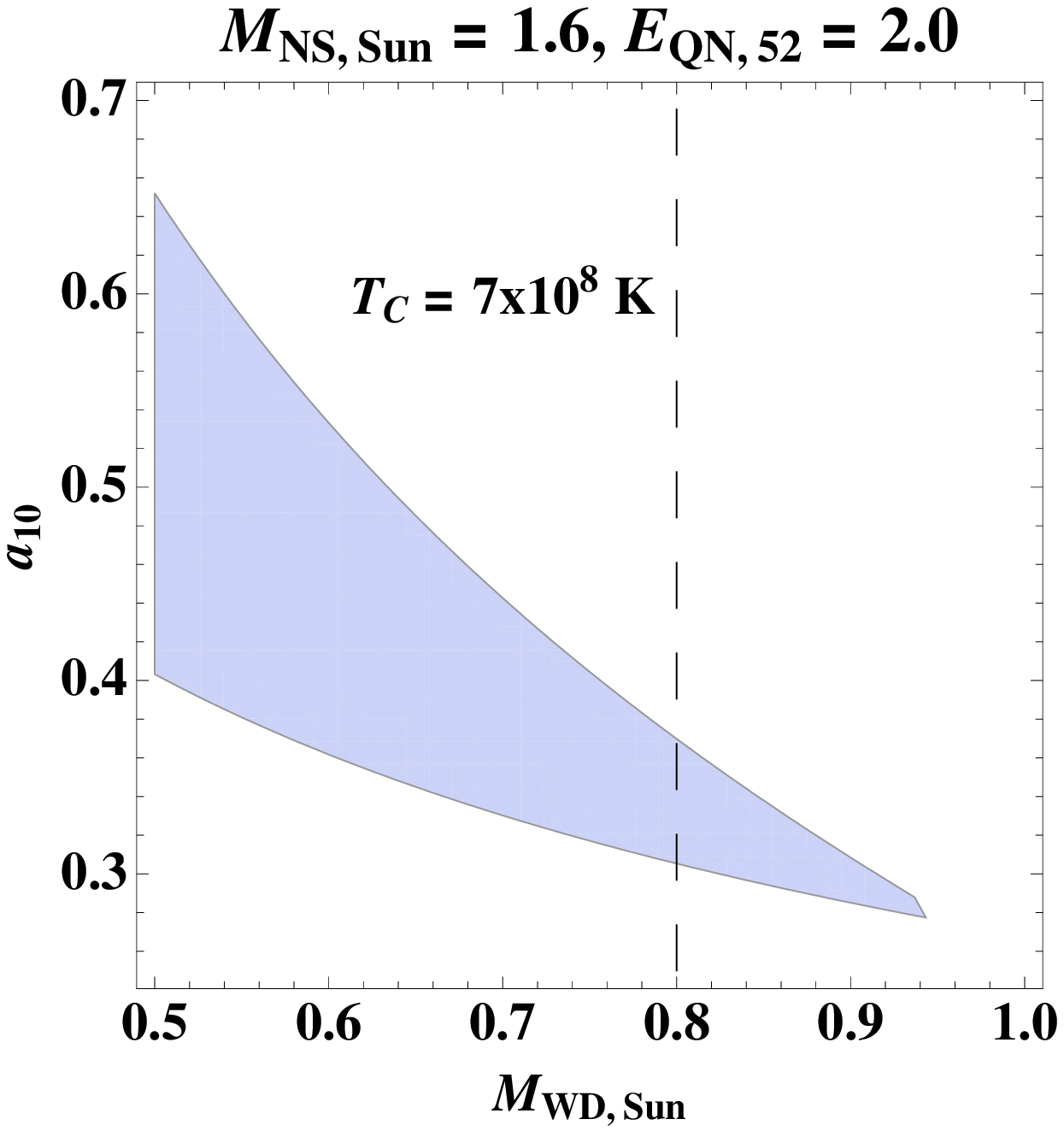} & \includegraphics[scale=0.43]{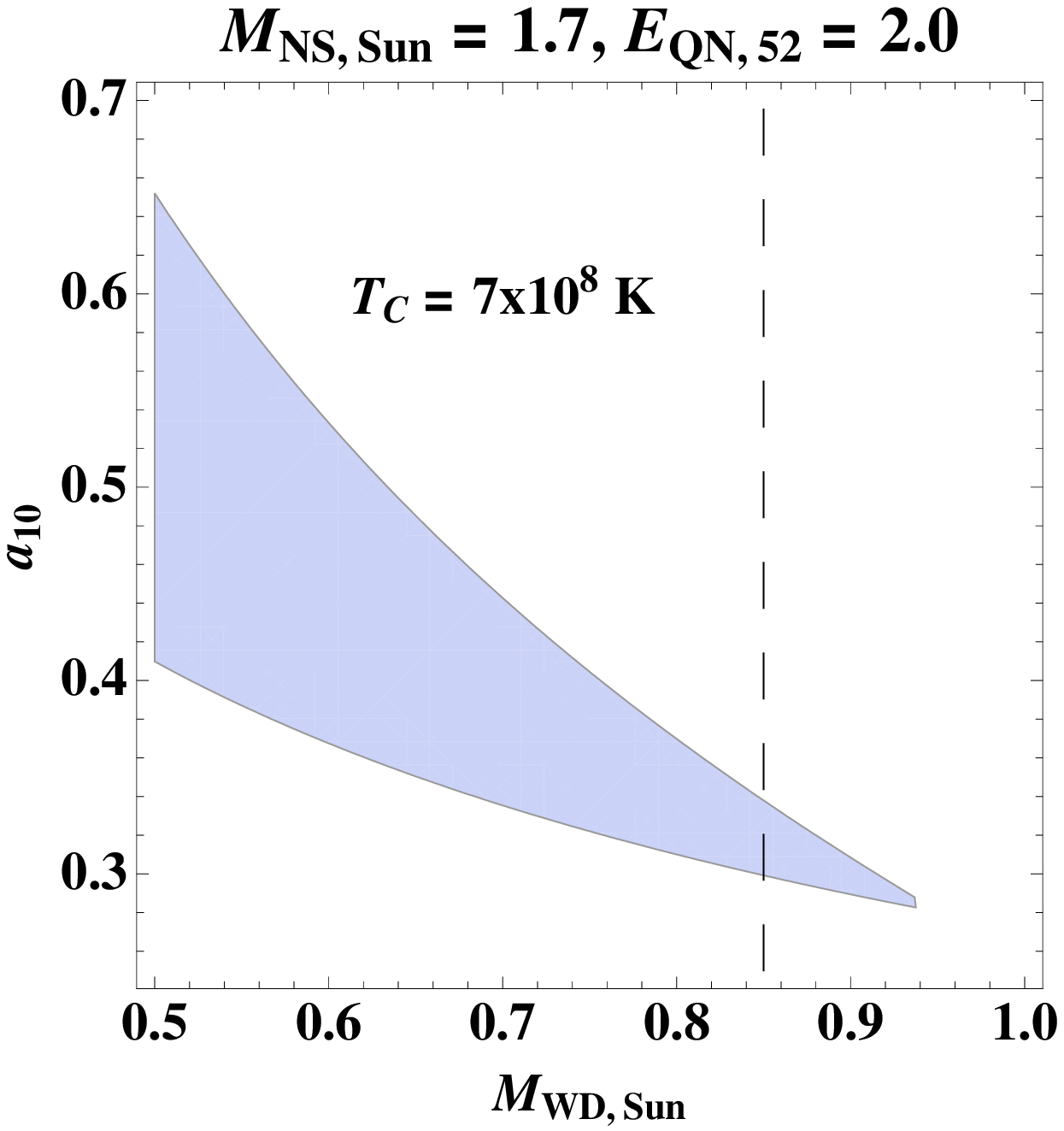} & \includegraphics[scale=0.45]{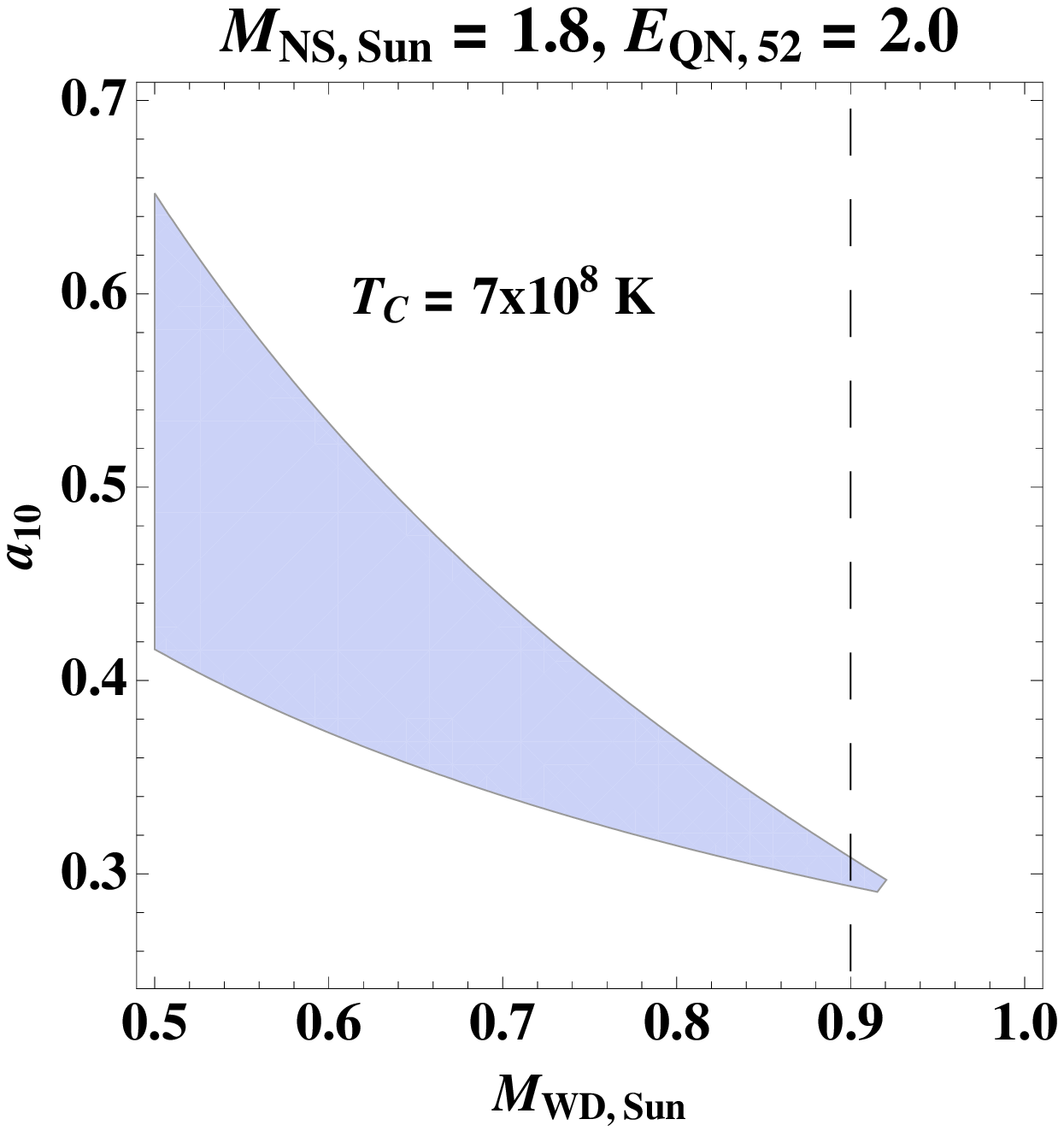} \\
\includegraphics[scale=0.43]{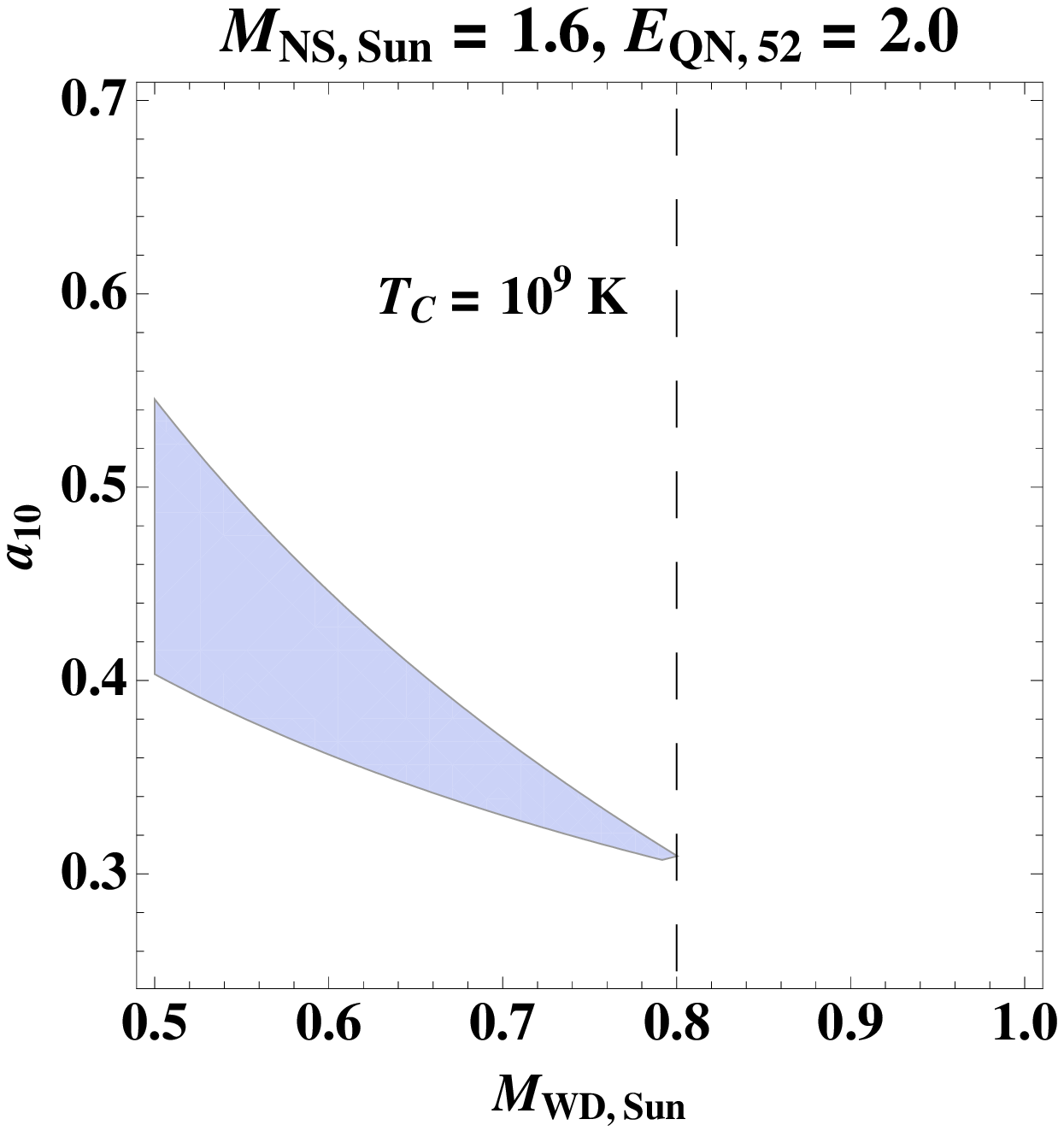} & \includegraphics[scale=0.43]{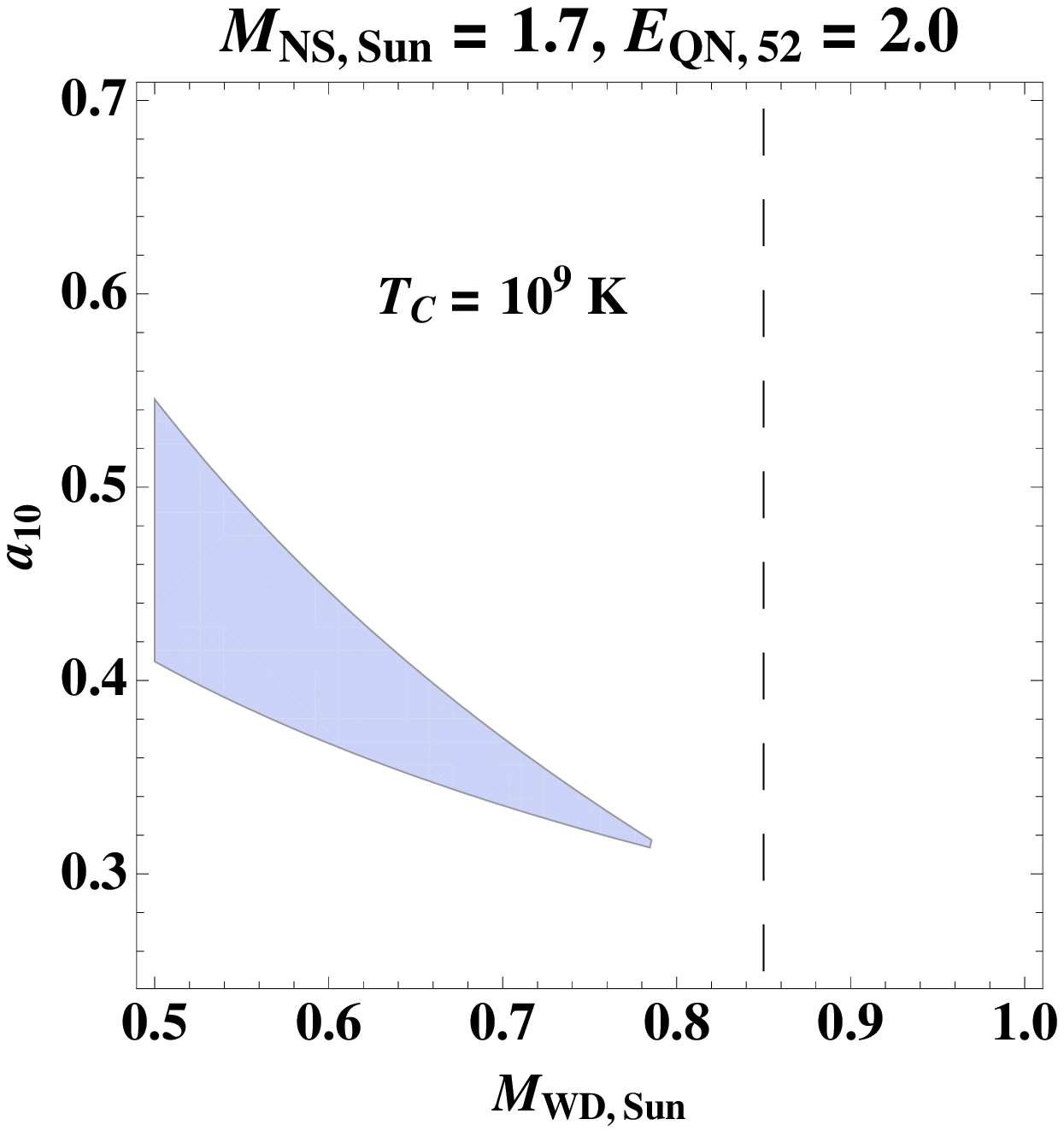} & \includegraphics[scale=0.45]{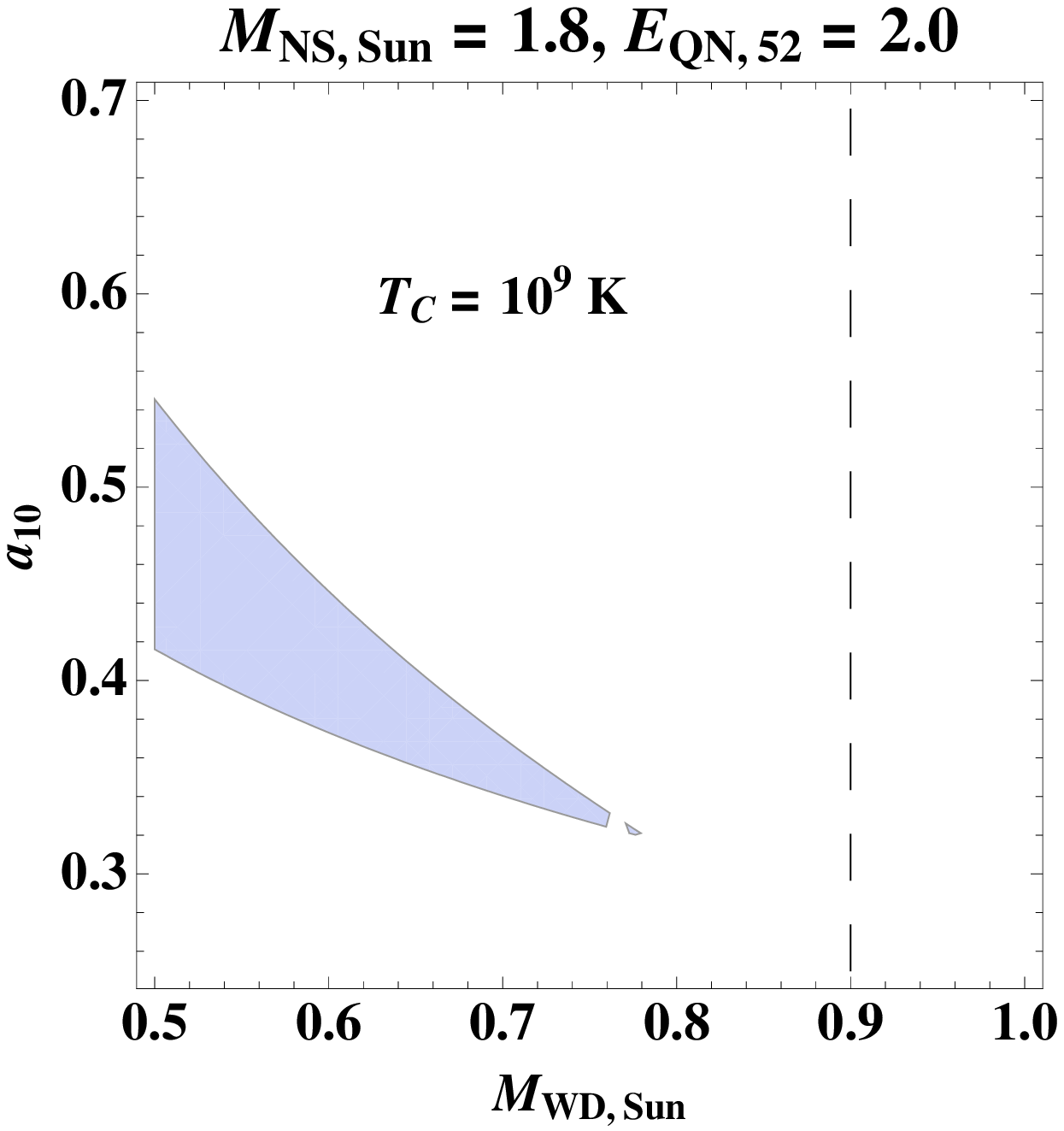} 
\end{tabular}
\caption{The  range in WD mass and orbital separation (in units of $10^{10}$ cm) that could result in 
 runaway C-burning after QN impact for $M_{\rm NS}=1.6M_{\odot}, 1.7M_{\odot}$ and 1.8$M_{\odot}$ from left
 to right, respectively.   {\bf Top panels}:  $T_{\rm C}=7\times 10^8$ K and $E_{\rm QN,52}^{KE}=1.0$. {\bf Middle panels}:  $T_{\rm C}=7\times 10^8$ K and $E_{\rm QN,52}^{KE}=2.0$. {\bf Bottom panels}:  $T_{\rm C}= 10^9$ K and $E_{\rm QN,52}^{KE}=2.0$.
Higher $T_{\rm C}$ requires slightly higher $E_{\rm QN}^{KE}$.
}
\label{fig:fig1}
\end{center}
\end{figure*}

The constraint above also guarantees that 
   the average temperature   of the shocked WD is
high enough to burn Carbon ($T_{\rm SWD} > T_{\rm C}\simeq 7\times 10^8$ K; e.g. Nomoto 1982). 
 The $T_{\rm SWD} > T_{\rm C}$ condition 
 puts a constraint on the   separation between the MNS and the COWD:
\begin{equation}
\label{eq:aC}
a_{9}  < a_{\rm C, 9} \simeq 3.7 \times {(E_{\rm QN, 52}^{\rm KE})}^{1/2}
\frac{\left(f(M_{\rm WD})/0.68\right)^{1/2} }{  \mu_{\rm WD, 2}^{7/6}  M_{\rm WD,0.6}^{5/6}}\ ,
\end{equation}
 which shows that  $a_{\rm C} > a_{\rm nuc}$ can be safely assumed given 
  the weak dependence of $a_{\rm nu.}$ on $\rho_{\rm SWD}$. 
 This allows to  effectively collapse eqs. (\ref{eq:anuc}) and (\ref{eq:aC})  to one  equation with $T_{\rm C}$ as the free common parameter.
   Hereafter we restrict ourselves to $a < a_{\rm C, 10}$. 
Furthermore, burning  occurs under degenerate 
conditions  since the Fermi temperature for the WD relativistic gas (e.g. Shapiro\&Teukolsky 1983; p24), $T_{\rm F} \sim 2\times 10^{10}\ {\rm K}\times (\rho/  10^8\ {\rm g\ cm}^{-3})^{1/3}$, guarantees
$T_{\rm F} > T_{\rm SWD} > T_{\rm C}$ in our model. The $T_{\rm F} > T_{\rm SWD}$
puts a lower limit on the WD mass;  in the helium WD regime which we do not
consider here.

The constraint on the binary separation $a <  a_{\rm C}$  hints at tight MNS-COWD  systems.
The tightest orbit  is reached when the WD overflows its Roche lobe (RL)   at 
   \begin{equation}
   \label{eq:aRL}
   a= a_{\rm RL} \simeq 2.2\times 10^{9}\ {\rm cm}\times  \frac{g(q)}{M_{\rm WD, 0.6}^{1/3}} \ ,
   \end{equation}
 where $g(q) = 0.6 + q^{-2/3} \ln{(1+q^{1/3})}$ (Eggleton 1983).  
 
 Figure \ref{fig:fig1} shows the range in WD mass which satisfies  $a_{\rm RL} \le a <  a_{\rm C}$ 
 when the QN goes off.  The range is $ 0.5M_{\odot} < M_{\rm WD} < 0.9M_{\odot}$ for our fiducial
 values.  The lower limit as we have said is because we only consider CO WDs  in a relativistic degenerate regime. 
 The upper limit $\sim 0.9M_{\odot}$, is restricted  by the maximum $E_{\rm QN}^{\rm KE}$ value adopted.
 We note that
  for high $T_{\rm C}$ values, solutions are found by increasing the kinetic energy of the QN ejecta.
  In Fig.  \ref{fig:fig1} ,  the $M_{\rm NS,c.}=1.6M_{\odot}$ lower NS mass limit corresponds to the BBB2 EoS which is the softest EoS we adopted in our calculations.
  The $M_{\rm NS,c.}=1.8M_{\odot}$ corresponds to the  APR EoS.
  The vertical dashed line shows the $q=0.5$ limit which we assume defines
   the non-merging (i.e. $q\le 0.5$) regime.
   
     The WD mass range in our model implies $q= M_{\rm WD}/M_{\rm NS}\le 0.5$ (see  figure \ref{fig:fig1} ) which  means that the binary system may not merge. Instead when the orbital separation will shrink below $a_{\rm RL}$ 
    accretion from the WD onto the NS ensues   which then  increases orbital
  separation above $a_{\rm RL}$.  The system eventually stabilizes itself   around $a_{\rm RL}$ (e.g. D'Souza et al. 2006)
    within the optimum orbital separation ($a_{\rm RL} < a < a_{\rm C}$) for 
   the WD to explode as a Type Ia when it is impacted by the QN shock\footnote{ The case where the QN goes off when $a< a_{\rm RL}$  requires a separate treatment (and will be explored elsewhere) since  one needs to take into account the WD radius increase following
    RL overflow and possible temporary lifting of degeneracy.}.  Recall that the NS would have been
     born very close to $M_{\rm NS,c}$ (see \S \ref{sec:Mc}), such that very little accretion can cause a QN (i.e.
     before much angular momentum has been accreted). 
    Ideally then is for the NS to explode when the system has settled into an $ a\sim a_{\rm RL}$ 
  configuration which points at specific QN-Ia progenitors  as discussed in \S \ref{sec:progenitor}.

   Our model is fundamentally different from other sub-Chandrasekhar models since
  the extreme compression by the QN shock creates conditions close
  to higher mass WDs with average densities exceeding $\sim 10^8$ g cm$^{-3}$ (see discussion in \S 5.1).
  It remains to prove that the high densities  and temperatures can successfully  ignite carbon
   which would require extensive numerical simulations to answer.  
   These would  include the details of shock compression, propagation and subsequent ignition.
    For now, we would argue     that our model
 possesses features that could lead to homogenous and efficient ignition in the core (or may be near
  the $\sim R_{\rm WD,c}$ region) of the shocked WD configuration.

  \subsection{Nuclear products}
  \label{sec:products}
           
           A distinctive feature of SNe-Ia spectra near maximum light is the presence of  
           intermediate-mass elements (IMEs) from Si to Ca, moving at velocities of
       10,000 to 16,000 km s$^{-1}$ (Pskovskii 1969; Branch et al. 1982, 1983; Khokhlov 1989; Gamezo et al. 1999; Sharpe 1999).     According to observations, $\sim$ 0.2-0.4$M_{\odot}$ of IMEs have to be synthesized during the explosion (see also Iwamoto et al. 1999).   
        Here we discuss the production of  Iron-Peak Elements (IPEs) and IMEs in our model while the resulting
       expansion velocities are discussed in \S \ref{sec:lightcurve&spectrum}.

        IPEs are produced in regions that reach $\sim 4\times 10^9$ K before degeneracy is lifted,
  which requires a density $\rho \ge 10^7$ gm cm$^{-3}$ (and $\rho \ge 2\times 10^6$ gm cm$^{-3}$ for IMEs;
   e.g. Woosley \& Weaver 1986).  For the same reason, ignition in single-degenerate sub-Chandrasekhar mass WD models require
   $M_{\rm WD} > 0.9M_{\odot}$; for  lower WD mass the $^{56}$Ni yield is tiny (Sim et al. 2010).  As noted earlier,
    the sub-Chandrasekhar mass WD mergers (van Kerkwijk et al. 2010) lead
   to a cold remnant ($\sim 6\times 10^8$ K) with central densities $\sim 2.5\times 10^6$ gm cm$^{-3}$. However, 
   accretion of the thick ``disk"  leads to compressional heating 
   resulting in an increase in the central temperature to $\sim 10^9$ K and densities $\sim 1.6\times 10^7$ gm cm$^{-3}$.
   These conditions they argue could  ignite the remnant centrally  with the nuclear runaway inevitable. 
    These centrally detonated WDs  look like ordinary SN-Ia (Sim et al. 2010; Kromer et al. 2010)
     and removes the need for a deflagration;   meaning that a deflagration is not necessary to produce the observed IMEs. 
     
    While compression and heating in the van Kerkwijk et al. (2010) is provided by accretion, 
     in our model it is provided by the QNE impact. If ignition occurs in (or close to) the center
      as we argued in \S \ref{sec:compression}, then in principle the resulting explosion should produce
       a composition relatively similar to observed ones. However, the very small WD mass in our model ($M_{\rm WD} < 0.9M_{\odot}$)
        and the extreme compressions from
       the QN shock, could mean that  more  IPEs are produced at the expense of IMEs. 
       
          It is hard to estimate the exact amount of IMEs produced in QNe-Ia explosions without detailed simulations but an order of magnitude
  estimate is arrived at as follows:  For our fiducial values, we estimate that only a small percentage ($<$ 10\%)  of the WD radius (or rather at most a few
  percents  of the WD mass) encloses densities less than $10^5$-$10^6$ g cm$^{-3}$ (e.g. Even\&Tohline 2009).  The shock will compress these layers to an average density of $\sim 10^7$ g cm$^{-3}$ where IMEs can eventually be produced.  
  Given these rough numbers above, it not unreasonable to assume
  that  typical QNe-Ia, at least for the fiducial values we chose, could convert up to 90\%  of the WD mass into  $^{56}$Ni (and thus IPEs) and at most 10\% into IMEs.
   I.e. that on average a typical QN-Ia involving the explosion of $\sim 0.6M_{\odot}$ WD would produce, under optimum
  conditions, up to  
      $\sim 0.45M_{\odot}$ of $^{56}$Ni. (i.e. of IPEs).
  Lower mass CO WDs, with a higher percentage of their mass at densities below $10^{6}$ g cm$^{-3}$ (Even\&Tohline 2009),   would be compressed less and should  produce more IMEs at the  expense of  $^{56}$Ni.  These numbers should serve as a very rough estimation
   of the average $^{56}$Ni yields in our model.  Quantitative evaluation of the $^{56}$Ni yields awaits detailed simulation.
   Hereafter we adopt $M_{\rm Ni}\sim 0.3M_{\odot}$
   as our fiducial value for the average $^{56}$Ni yield in a typical QN-Ia.

 In the standard models of SNe-Ia, the diversity of SNe-Ia reflected in the range of peak luminosity provides a direct measure of the mass of $^{56}$Ni ejected/synthesized
      varying from  $\sim 0.1 M_{\odot}$ associated with the sub-luminous objects to $\sim 1.3M_{\odot}$ for the most luminous events (e.g. Stritzinger et al. 2006a; Wang et al. 2008 to cite only a few).  A number of potential parameters could influence the amount of $^{56}$Ni  produced, 
      e.g., C-O ratio, overall metallicity, central density,
      the ignition intensity, the number of ignition points in the center of WDs or the transition density from deflagration to detonation
      as well as asymmetry in the explosion (e.g. Timmes et al. 2003; R\"opke \& Hillebrandt 2004; R\"opke et al. 2005; Stritzinger et al. 2006b; Lesaffre et al. 2006; Podsiadlowski et al. 2008; Kasen et al. 2009; H\"oflich et al. 2010;
      Meng et al. 2011). Nevertheless, it seems that 
      the origin of the variation of the amount of $^{56}$Ni  for different SNe-Ia is still unclear.

     Our model  might provide a range in  $^{56}$Ni mass  by a one parameter sequence in terms of the WD mass, 
if all QNe-Ia were to occur  (or triggered) when $a=a_{\rm RL}$.   Other fiducial parameters such as $\Gamma_{\rm QNE}$ do not vary much from one system to another
          since the condition for QN explosion is a universal one defined by the quark deconfinement density (here
          chosen to be $5\rho_{\rm N}$). However,    the upper COWD mass in our model
           ($\sim 0.9M_{\odot}$) means that to account for the extreme
            (up to $\sim 1.3M_{\odot}$) masses of $^{56}$Ni observed  one might require the SD, the sub-Chandrasekhar WD mergers,  and/or
            DD channel.  As we show below, spin-down power from the QS
             could brighten the explosion mimicking standard SNe-Ia with $M_{\rm Ni}> 1 M_{\odot}$.
              In other words, these spin-down powered QNe-Ia  could be mis-interpreted
                as standard SNe-Ia with much higher  $^{56}$Ni  content 
              than truly processed/produced.

     \section{The light-curve and the spectrum}
     \label{sec:lightcurve&spectrum}
     
      \subsection{The bolometric light-curve}
  \label{sec:lightcurve}

In standard SNe-Ia, the ejecta remain optically thick for the first several months after explosion. The width of the bolometric light curve is related to the photon diffusion time, $\tau_{\rm d}$. The peak of the light-curve is directly related to the mass of $^{56}$Ni produced by the explosion.
 If powered  by $^{56}$Ni decay alone (hereafter $^{56}$Ni-powered meaning powered by $^{56}$Ni and $^{56}$Co decay), QNe should produce light-curves
  that most likely obey the Phillips relationship (i.e. in the B-band). On average, QNe-Ia  
              should appear   less broad and dimmer (with $M_{\rm Ni} < 0.45M_{\odot}$; we adopt an average of $M_{\rm Ni}\sim 0.3M_{\odot}$) than their Chandrasekhar mass counterparts (with $M_{\rm Ni}\sim 0.6M_{\odot}$).  
                However,   there are reasons to expect QNe-Ia to be brighter with broader light-curves than standard SNe-Ia of
  similar  $^{56}$Ni yields if a mildly or a rapidly rotating QS is left behind by the QN.        
             The energy deposited into the expanding WD ejecta by the spinning-down QS can substantially brighten the light-curve.
              Or at least it can compete with the decay of $^{56}$Ni and thermal energy in the expanding WD material.
              This is reminiscent of powering of Type II SNe shell by pulsar spin-down. 
              Maeda et al. (2007) proposed that some ultraluminous supernovae may be explained by dipole emission from a rapidly spinning magnetar, which was worked out in detail by Kasen \& Bildsten (2010) and Woosley (2010). 
               In the early stages, the majority of the spin-down energy is  lost to adiabatic expansion and not seen directly in the peak luminosity. 
                      Eventually, a percentage of the energy goes into kinetic energy of expansion, while  the
                    remainder goes  into radiation.         
              Depending on the magnetic field strength and the NS's initial period,
                 these studies find that  a percentage (up to $\sim$ 50\%) of the spin-down energy went into kinetic energy of expansion, while the remainder went into radiation.      In certain cases, the  spin-down powered ejecta could lead to a 
                  light-curve peak of  about a few times the peak luminosity of typical Type Ia supernovae. 
                  Furthermore, the resulting light-curves are broader than without the spin-down power injection.

                            As in these models,     the spin-down energy in QNe-Ia should result in an additional entropy injection
               (on timescales exceeding the adiabatic expansion phase) that should  brighten the
                light-curve; and may be  modify its shape.    
                The energy can accelerate the WD ejecta to slightly higher ``coasting" velocities  than in standard (purely $^{56}$Ni-powered) SNe-Ia.
                The tail of the light-curve could resemble radioactive decay for some time but, assuming complete trapping of the spin-down emission, would eventually be brighter (e.g. Woosley 2010).  In other words, compared to light-curves from
purely $^{56}$Ni-powered light-curves, QNe-Ia light-curves should  show some differences in the
rise and fall time.  From these considerations, we are tempted to argue that QNe-Ia light-curves should not  obey the Phillips relationship. 
Although the light-curve in the B-band would need to be computed to corroborate this point.

\begin{figure*}
\begin{center}
\begin{tabular}{ccc}
\includegraphics[width=0.5\textwidth]{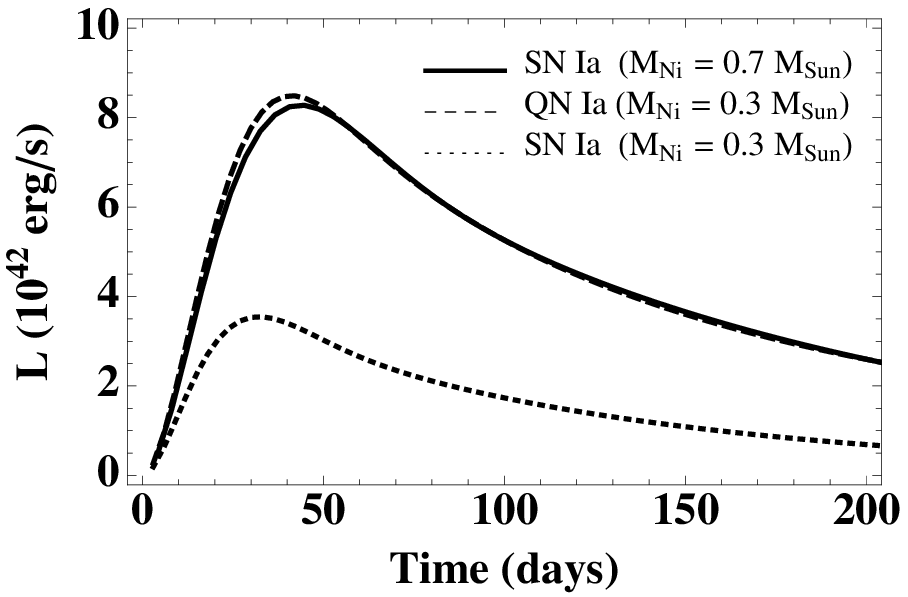}
\includegraphics[width=0.5\textwidth]{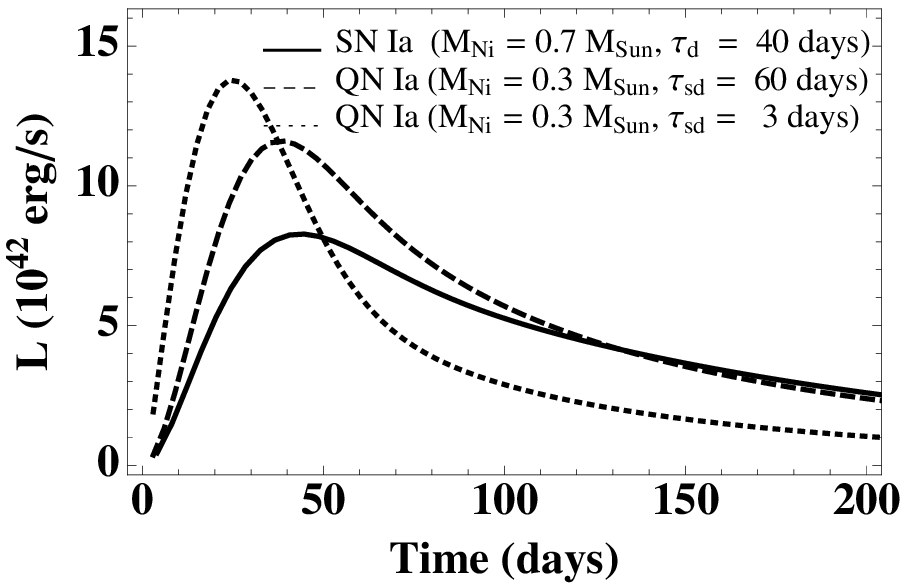}
\end{tabular}
\caption{The solid line in the two panels shows the bolometric light-curve for $^{56}$Ni and $^{56}$Co decay in
 a standard SN-Ia model with  $M_{\rm Ni}=0.7M_{\odot}$ and diffusion time $\tau_{\rm d}= 40$ days.
{\bf Left panel}:  The dashed  curve shows a $M_{\rm Ni}=0.3M_{\odot}$ QN-Ia  bolometric  light-curve with  spin-down energy 
 $E_{\rm sd}=2\times 10^{50}$ ergs (i.e.  $P_{\rm QS}=10$ ms)  and  $\tau_{\rm sd}=4.5\tau_{\rm d}$ with  $\tau_{\rm d}= 30$ days.  
 For comparison a   purely $^{56}$Ni-powered SN-Ia light-curve   with  $M_{\rm Ni}=0.3M_{\odot}$ and diffusion time $\tau_{\rm d}= 30$ days is also shown (dotted line).
 The choices or the parameters for the $M_{\rm Ni}=0.7M_{\odot}$ standard SN-Ia (solid curve) were made such that it shows
 close overlap with the spin-down powered QN-Ia. 
 {\bf Right panel}:   The dashed line is a typical QN-Ia in our model (i.e. $M_{\rm Ni}=0.3M_{\odot}$, $E_{\rm sd}=2\times 10^{50}$ erg and $\zeta_{\rm sd}=0.5$) with $\tau_{\rm sd}= 60$ days. The dotted line is the same QN-Ia but for $\tau_{\rm sd}= 3$ days.
 For details on the models see \S \ref{sec:lightcurve}.
}
\label{fig:fig2}
\end{center}
\end{figure*}

                   No simple analytic solution for the bolometric light-curve exists when taking into account $^{56}$Ni-decay
   power and spin-down power. Instead we make use of semi-analytical models presented in
    Chatzopoulos et al (2012)  to illustrate our point (assuming that the spin-down energy
     is thermalized throughout the expanding WD material). 
         In the left panel in Figure \ref{fig:fig2} we compare a standard purely $^{56}$Ni-powered SN-Ia with  $M_{\rm Ni}=0.3M_{\odot}$
       and diffusion time $\tau_{\rm d}=30$ days (the dotted line) to a QN-Ia (the dashed line) with  $M_{\rm Ni}=0.3M_{\odot}$ 
      but boosted by spin-down energy of $2\times 10^{50}$ erg 
    (i.e.  $P_{\rm QS}=10$ ms) and $\tau_{\rm sd}= 4.5\tau_{\rm d}$ (i.e. $B_{\rm QS}\simeq 9\times 10^{13}$ G); 
     we keep  $\zeta_{\rm sd}=0.5$ (i.e. half of the spin-down energy went into radiation).
    This illustration, together with what has been  inferred from the above mentioned studies, suggest that    
     QNe-Ia should be brighter and broader than their purely $^{56}$Ni-powered counterparts of similar $^{56}$Ni yield.
    For a comparison we also show  a purely $^{56}$Ni-powered SN-Ia with  $M_{\rm Ni}=0.7M_{\odot}$
                   and $\tau_{\rm d}=40$ days. It was chosen so that it closely overlaps with the QN-Ia light-curve.
                   It is shown as the solid line in both panels in Figure \ref{fig:fig2}.
      
      To demonstrate the effect of the spin-down timescale on the light-curves,  
                                  the right panel in Figure \ref{fig:fig2}  compares our chosen  standard SN-Ia to a typical
                                   QN-Ia (i.e. $M_{\rm Ni}=0.3M_{\odot}$, $E_{\rm sd}=2\times 10^{50}$ erg and $\zeta_{\rm sd}=0.5$) 
                                   but for different  $\tau_{\rm sd}$; 
                  $\tau_{\rm sd}= 60$ days (the dashed line) and the other $\tau_{\rm sd}= 3$ days (the dotted line). 
                  While the $\tau_{\rm sd}= 3$ days QN-Ia is narrower, a slight increase in its $^{56}$Ni content will
                  make it broader and brighter than the standard SN-Ia. We find that 
                  although QNe-Ia are expected to  have somewhat distinct light-curves, some would appear relatively similar to standardizable SNe-Ia. 
For example, and in general,  we find that for $0.2 \le M_{\rm Ni}/M_{\odot} \le 0.4$ and $0.2 \le \zeta_{\rm sd} \le 0.4$
                   a range in QNe-Ia  light-curves  can be found that closely overlap our standard  SN-Ia. 
This shows that based on (bolometric) light-curves alone, QNe-Ia could be mistaken for standard SNe-Ia.      
But in general QNe-Ia should deviate from the Phillips relationship.  
                The implications  to cosmology will be discussed in \S \ref{sec:cosmology}.

                  The peak of the light-curve in a spin-down powered QN-Ia   would not be directly related to its mass (i.e. the amount of $^{56}$Ni
  produced) but would  also be sensitive to the dipole field strength and the initial period of the QS (i.e. $B_{\rm QS}$ and $P_{\rm QS}$). 
    If QNe-Ia exist, and mistaken for
                standard SNe-Ia, this additional energy input
                would lead to an overestimate of the amount of $^{56}$Ni produced by the explosion when using Arnett's law (Arnett 1982).
                               Following Stritzinger \& Leibundgut (2005), we can write Arnett's law as   
               \begin{equation}
               L_{\rm Ni} (t_{\rm r}) = \alpha \ ( 6.45 e^{- t_{\rm r}/8.8 {\rm d}} + 1.45e^{- t_{\rm r}/111.3 {\rm d}} )  \times 
                 \frac{M_{\rm Ni}}{M_{\odot}}  \times 10^{43}\ {\rm erg\ s}^{-1}\ , 
                   \end{equation}     
                where $\alpha$ is a correction factor of order unity to ArnettÕs law and $t_{\rm r}$ is the time between explosion and maximum light (i.e., the bolometric rise time).   For $\alpha=1$ and assuming a typical, normal SN-Ia to have   a rise time of $\sim 18$ days (e.g. Hayden et al. 2010; Phillips 2012),
                we get      
               $L_{\rm Ni} (18d) \sim 2.1\times 10^{43}\ {\rm erg\ s}^{-1}\times  M_{\rm Ni}/M_{\odot}$ while  $L_{\rm sd}(18d)\sim  2.4\times 10^{43}\ {\rm erg\ s}^{-1} \times\zeta_{\rm sd,0.5}/(B_{\rm QS, 15}^2 P_{\rm QS, 10})^{2/3}$ where $\zeta_{\rm sd, 0.5}$ is the percentage of spin-down energy that went into radiation in units  of 0.5.  
               Since  $L_{\rm Ni} (18d)\sim L_{\rm sd} (18d)$ for our fiducial values (i.e. a total peak luminosity of $L_{\rm tot.} \sim 2\times L_{\rm Ni}(18d)$)   
                   this would overestimate the
                 average $^{56}$Ni yield  by a factor of up to a few.  This would appear troubling since
                  the spectrum would probably be indicative of a much lower  $^{56}$Ni content (see \S \ref{sec:candidates}). 
                  The spin-down energy  in our model is unique and it effectively
     separates the photometry from the spectroscopy.

 \subsection{The spectrum}
 \label{sec:spectrum}

             The double-peaked structure observed in the NIR light-curves of typical SNe-Ia is a direct sign of the concentration of IPEs\footnote{The secondary maximum often found in R and I (and more prominently in the NIR) is attributed to the cooling of the ejecta to temperatures where the transition from Fe III to Fe II becomes favorable, redistributing flux from shorter wavelengths to longer wavelengths (H\"oflich, Khokhlov, \& Wheeler 1995; Kasen 2006). } in the central regions, whereas the lack of a secondary maximum is indicative of strong mixing.
                  Specifically, from models of the radiative transfer within SNe, Kasen (2006) finds that the timing and strength of the shoulder is dependent on the distribution and amount of  $^{56}$Ni within the ejecta. Models with a completely homogenized composition and with a small amount of $^{56}$Ni result in an $i$-band light-curve with no discernible secondary peak or shoulder. Instead, the two peaks merge to produce a single broad peak.

                               QNe-Ia produce less $^{56}$Ni than standard luminous SNe-Ia. Thus, inherently,  QNe-Ia   are expected to be lacking a (or  showing a weak)  secondary maximum.    If spin-down energy is negligible (or the energy is deposited in a jet-like structure
 away from the viewer's line of sight)\footnote{In the QN-Ia model, 
                    the spin-down source is offset from the WD explosion point by a distance of the order of $a_{\rm RL}$ (i.e.
                  the binary separation). However, it takes only a few seconds for the WD expanding ejecta to engulf the QS.
                   How the spin-down energy is deposited (isotropically or with a jet-like structure) and how
              it is dissipated in the WD material remains to be investigated and is beyond the scope of this paper.
},   we expect  spectra indicative of unmixed burning with the  radioactive Nickel (and thus IPEs) produced mainly in the denser core ($> 10^8$ g cm$^{-3}$) and the IMEs in the outer layers (at higher expansion velocities).  These should appear in many ways (spectral
 features and light-curves) similar to 
   sub-luminous SNe-Ia and should obey the Phillips relationship. However, the lack of a prominent second maximum in the $i$-band
    should distinguish a QN-Ia from a standard SN-Ia.

     On the other hand,  taking into account spin-down energy,
 the central overpressure caused by the energy deposition from spin-down should  blow a bubble in the expanding WD material, similar to the dynamics studied in the context of pulsar wind nebulae (e.g., Chevalier \& Fransson 1992). As shown in multi-dimensional calculations of pulsar wind nebulae, as the bubble expands, Rayleigh-Taylor instabilities would mix the swept-up material (e.g. Blondin et al. 2001).  This could
             in principle occur in our case which should result in the dredging-up of burnt core material (IPEs) to the surface and 
             IMEs to the core.   This mixing  should manifest itself in       the presence of IPEs at higher velocity than the IMEs. 
                    Besides suppressing the $i$-band shoulder
                   (i.e. secondary maximum),  mixing will also have important implications for the spectrum, especially at late times
                    and may affect the amount of IPEs and IMEs processed during the expansion.
                 A more detailed analysis would require 
 multi-dimensional studies of the coupled radiation transport and hydrodynamics, but are postponed for now.

\subsection{Plausible QNe-Ia candidates}
\label{sec:candidates}

  Observationally, SNe-Ia have been classified into three
subclasses: normal SNe-Ia, overluminous SNe (SN 1991T-like), and faint SNe-Ia (SN 1991bg-like) (Branch et al. 1993; Filippenko 1997; Li et al. 2001). The light-curves of more luminous SNe-Ia decline more slowly (Phillips 1993).
  More recently, a range of properties for peculiar subluminous SNe-Ia has been discovered:

  \begin{itemize}
  
  \item {\it SN2006bt-like objects (braod light-curves but spectroscopically subluminous)}: Foley et al. (2010) presented evidence that SN 2006bt spectroscopically resembled SN 1991bg (subluminous, fast declining SN-Ia), but photometrically resembled a normal SN-Ia.  I.e. it has a slowly-declining light-curve characteristic of luminous SN-Ia but with spectra displaying absorption features characteristic of low-luminosity SN-Ia.
    Maguire et al. (2011) presented data on the subluminous PTF 10ops which shared many similarities with SN 2006bt. PTF 10ops also had a broad light-curve.  SN2006ot, appears to be related to SN 2006bt (Stritzinger et al. 2011). The photometry for this object shows that the two SNe were quite similar. This similarity also extended to the peak absolute magnitudes, which were the same to within $\sim  0.1$ mag: a broad light-curve characteristic of luminous SN-Ia ($\Delta m_{15}$(B) = 0.84 mag), but a weak secondary $i$-band maximum characteristic of low-luminosity events. Spectroscopically, however, SN 2006ot showed  differences with respect to SN 2006bt although Stritzinger et al. (2011) find that at 3-4 weeks past maximum light, the spectra of SN 2006ot are similar  stressing the similarities between these two SNe.
    
  \item  {\it SN2003fg-like objects  (overluminous but spectroscopically subluminous)}:  SN-Ia 2003fg 
  is an extremely luminous SN-Ia. Howell et al. (2006) have  concluded that SN 2003fg is very likely a super-Chandrasekhar mass SN-Ia perhaps with a mass $\sim 2 M_{\odot}$. Other similar object  include 2006gz (Hicken et al. 2007) and SN 2007if (Akerlof et al. 2007).
 Despite the extreme luminosities, these SNe show the slowest luminosity evolution (i.e. low velocities  of the expanding SN materials as deduced from  the spectra).
 The low velocity and short time-scale seen in SN 2003fg indicate that the ejecta mass is smaller than the Chandrasekhar mass, which is an apparent contradiction to the large luminosity.

Maeda\&Iwamoto (2009) noted that  these candidate over-luminous SNe-Ia 2003fg, 2006gz, and (moderately over-luminous) SN 1991T, have very different observational features: the characteristic time-scale and velocity are very different.  
In analyzing SN2003fg,  Maeda\&Iwamoto (2009) concluded that SN2003fg requires that either $M_{\rm Ni}$ or $M_{\rm WD}$  (or both) should be smaller than even the Chandrasekhar mass, contrary to the earlier expectations (Howell et al. 2006). On the other hand, the large peak luminosity requires that  $M_{\rm Ni} \sim 1.1M_{\odot}$.   They also concluded that the observed features of SN 2006gz are consistent with expectations from the super-Chandrasekhar mass WD explosion scenario. They 
 suggest that the observed differences can be attributed to different viewing orientations if the progenitor WD, and thus the SN explosion, is aspherical.

   \item {\it SN2002cx-like objects (subluminous but spectroscopically overluminous)}:  These objects have maximum-light spectra similar to those of overluminous objects like SN 1991T. However, the expansion velocities of these objects at maximum light indicate an explosion with low kinetic energy per unit mass (i.e.
   subluminous; Filippenko 2003; Li et al. 2003).  For example, SN2002cx had 
   expansion velocities approximately half those of ordinary SNe-Ia. The peak absolute magnitudes in B and V were nearly 2 mag fainter than a normal SN-Ia of the same decline rate, and the $i$-band light-curve displayed a broad primary maximum completely lacking a secondary maximum.
   SN2002cx-like objects distinguishing properties include: low luminosity for their light-curve shapes, a lack of a second maximum in the NIR bands, low photospheric velocities,  and a host-galaxy morphology distribution highly skewed to late-type galaxies (Foley et al. 2009; Valenti et al. 2009).
   In general, there appears to be a great diversity among SN 2002cx-like objects, with a distribution of absolute luminosity and kinetic energy (McClelland et al. 2010).

    \end{itemize}

\begin{figure}
\begin{center}
\begin{tabular}{ccc}
\includegraphics[width=0.4\textwidth,angle=90]{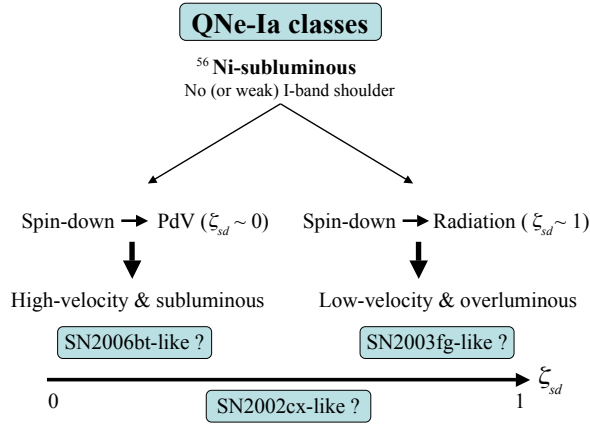}
\end{tabular}
\caption{The plausible manifestations (and tentative classification) of QNe-Ia ranging from   $\zeta_{\rm sd}\sim 0$ (i.e. most of spin-down power 
is used during the adiabatic expansion phase; i.e. into PdV work) to  $\zeta_{\rm sd}\sim 1.0$ (i.e. most of spin-down power went into radiation). We speculate that, and tentatively classify,   SN2006bt-like SNe as $\zeta_{\rm sd}\sim 0$ QNe-Ia,
 SN2003fg-like SNe as $\zeta_{\rm sd}\sim 1$  QNe-Ia. 
The  great diversity among SN 2002cx-like objects (with a distribution of absolute luminosity and kinetic energy; e.g. McClelland et al. 2010) 
we suggest reflects the range in $\zeta_{\rm sd}$  which can vary from 0 to 1 in our model.
}
\label{fig:fig3}
\end{center}
\end{figure}

  In a very general sense, and as summarized in Figure \ref{fig:fig3}, the composition, structure, and the energetics expected of  spin-down powered QNe-Ia  seem to resemble
   those  inferred for  the peculiar SNe-Ia objects (SN 2006bt-like, SN2003fg-like and may be SN2002cx-like). If the 
   spin-down energy is deposited anisotropicaly or in a jet-like structure than QNe-Ia observed off-axis
   may appear as normal SNe-Ia with low $^{56}$Ni content.
   
 Because of the spin-down energy, QNe-Ia's  photometric and spectroscopic properties  are not necessarily linked to each other.
  In QNe-Ia the spectrum is indicative of the amount
  of $^{56}$Ni produced  while the morphology and energetics of the light-curve can be affected (and probably dominated) by spin-down
  power.  Depending on the initial spin-down  energy (i.e. $B_{\rm QS}$ and $P_{\rm QS}$)
                   and $\zeta_{\rm sd}$ (the percentage of the spin-down energy that went to radiation),  low (high) velocities could accompany an  
                   over- luminous (sub-luminous) QN-Ia light-curve (see Fig. \ref{fig:fig3}).  The estimates of the  photospheric velocity at the maximum brightness 
                     and of the time-scale of the light-curve evolution around peak  are complicated in our model.
   The peculiar classes show another interesting feature:  unlike standard SNe-Ia with similar decline rates, they
  seem to be lacking  a prominent second maximum in the $i$-band; the ejecta 
  in these objects seem to be well mixed.  The low $^{56}$Ni yields in QNe-Ia and the
  efficient mixing likely to be induced by the pulsar-wind bubble provide conditions to erase (or at least minimize) the  $i$-band shoulder.

   \subsection{Summary}

                    The exact shape of a QN-Ia light-curve taking into account the spin-down power  remains to 
  be computed (in particular in the B-band) for a more robust comparison to a standard  light-curve. 
  For now we would argue that  spin-down powered QNe-Ia should be associated with fairly broad light-curves 
   with rise and decay time phases that should somewhat deviate from $^{56}$Ni-powered
   (i.e. standard) light-curves (see Figure \ref{fig:fig2}).  However while 
   QNe-Ia  would be associated with  somewhat distinct  light-curves, some would   still appear similar to  standard
   ones.    In the QN-Ia model, 
the additional (spin-down) power source effectively separates  photometry from spectroscopy.
 The $M_{\rm Ni}$ yield and the kinetic energy are not necessarily linked which means that the expansion velocities
  in QNe-Ia are not indicative of how powerful the explosion is.

 \section{The binary progenitor}
 \label{sec:progenitor}

  We ask what progenitor  could lead to tight  MNS-COWD system
  experiencing the QN event after it has settled into $a\sim a_{\rm RL}$?
   In others words a necessary condition is for $\tau_{\rm QN}$ to exceed
   the time it takes the MNS-COWD, from the moment of its birth,  to shrink its orbit to $a=a_{\rm RL}$.
     A non-accreting   NS-WD  system born with an initial
    period $P_0$, sees its orbit shrink due to gravitational wave (GW) emission.
The  orbital  period decay rate, neglecting orbital eccentricity, is
$ dP/dt  = -1.03\times 10^{-7}\ {\rm s~ s}^{-1}\  \left( 2\pi/P \right)^{5/3}\times (M_{\rm NS, \odot}M_{\rm WD,0.6})/M_{\rm T, \odot}^{1/3}$  (Landau \& Lifschitz 1975).
 A solution is $P (t)= P_0\times (1- t/\tau_{\rm GW})^{3/8}$ with
 \begin{equation}
 \label{eq:tGW}
 \tau_{\rm GW}\sim  10^7\ {\rm yrs}\ \frac{P_{\rm 0, hr}^{8/3}M_{\rm T, \odot}^{1/3}}{M_{\rm NS, \odot}M_{\rm WD,0.6}} \ ,
 \end{equation}
  with $P_{\rm 0, hr}$  the initial 
 period (in hours).  Here $M_{\rm T, \odot}$ is the total mass in solar units.
   
Any  interacting binary (with a primary $M_{\rm prim.}$ and a secondary $M_{\rm sec.}$) that 
     leads to an MNS-COWD system with $M_{\rm NS}\sim M_{\rm NS,c.}$ and
      a WD that has filled its RL is a potential candidate.   The accretion from the WD
      onto the NS will  eventually drive the NS above the $5\rho_{\rm N}$ triggering a QN event.
      However, there is also the possibility of the NS born with   $M_{\rm NS}> M_{\rm NS,c.}$
       but in a  fully recycled state ($P_{\rm NS} < 2$ ms) so that its core density is below
        the critical value. This case which we refer to as scenario 1 (hereafter S1) is discussed
        first  then followed by scenario 2  for mildly recycled case ($10\ {\rm ms} < P_{\rm NS} < 100$ ms; hereafter S2). 
        S1 and S2 involve a direct formation of the NS by iron-core collapse with one of the binary 
         component massive enough to lead to a MNS.  There is also an indirect path to
          forming a NS in a binary which appeals to the AIC of a WD to a NS (we will
          refer to this as S3; scenario 3). Mass transfer can drive a WD in a binary over the Chandrasekhar limit, which may lead to  an AIC (in the case of an O-Ne-Mg WD; and possibly also in some CO white dwarfs) which produces a NS. 
         This is  the most interesting one if it can be shown that a massive enough NS can result from the AIC of the WD.

 The mass of the progenitor star that could lead to a MNS (prone to the QN transition) 
  is in the 20-40$M_{\odot}$ range. For this range, 
     the QNe rate is estimated to be $\eta_{\rm QN} \le 1/100$ core-collapse (CC) events  ($\eta_{\rm QN}< 0.01 \eta_{\rm CC}$; Jaikumar et al. 2007; Leahy\&Ouyed 2008; Leahy\&Ouyed 2009; Ouyed et al. 2009). Assuming that 1/10 occur in tight binaries as required here this means
      a QN-Ia rate $\leq 1/1000$ core-collapse events in this scenario. This is lower than the
       currently observed Type Ia of $\sim 1/10$ of  core-collapse SNe (e.g.  Pritchet et al. 2008 and references therein).
       Unless the NS gain mass during the binary evolution towards the tight MNS-COWD system,
  both S1 and S2 will be plagued by low statistics (i.e.  direct dependency on $\eta_{\rm QN}$).       
       However, a top-heavy IMF of Pop. II stars together with   a boost in star formation rate at early times
        could make QNe-Ia from S2 statistically significant at high redshift.
        The  AIC channel provides better statistics provided enough of them lead to NSs which undergo a QN explosion.
                 Below we briefly describe each scenario.

 \subsection{MNS-COWD systems with fully recycled NSs}
 \label{sec:S1}

   The need for a COWD  in our model    and the requirement of a  fully recycled NS 
    hint at  IMXBs: (i) with a donor star in the $3.5 M_{\odot}< M_{\rm don.} < 5M_{\odot}$  range;
     (ii)  and have undergone a substantial accretion phase  (e.g.  Tauris et al. 2000). 
    In particular,  IMXBs evolving via the so-called ``Case A Roche-Lobe Overflow (RLO)" phase\footnote{There are three  types of RLO (e.g. Tauris et al. 2000; Podsiadlowski et al. 2002): cases A, B and C. In case A, the system is so close that the donor star begins to fill its Roche-lobe during core-hydrogen burning: in case B the primary begins to fill its Roche-lobe after the end of core-hydrogen burning but before helium ignition: in case C it overflows its Roche-lobe during helium shell burning or beyond. Cases B and C occur over a wide range of radii (orbital periods); case C even up to orbital periods of about 10 years. The precise orbital period ranges for cases A, B, and C depend on the initial donor star mass and on the mass ratio (see Tauris 2011). Once the RLO has started it continues until the donor lost its hydrogen-rich envelope (typically $>$ 70\% of its total mass) and subsequently no longer fills its Roche-lobe.},
      that evades spiral-in,  go through a mass-transfer phase lasting about $10^7$ yr.  These 
  lead to   mainly  millisecond pulsars (MSPs) with CO WD companions. 
  This evolutionary path provides enough material
   to    spin-up  a slowly rotating  NS to a MSP. Furthermore,  the  magnetic field of the MSP
would have decayed  in the process due to accretion (Taam \& van den Heuvel 1986).
Assuming the MSPs were born   with an initial magnetic field of $\sim 10^{12}$ G,
 accretion would decrease its surface magnetic field to $\sim 10^9$ G ($\sim 10^8$ G)
 by accreting only a few hundredth to a few tenths of $M_{\odot}$.
  This means that $\tau_{\rm QN} < 10^8$ ($<10^{10}$) years   with unbracketed and bracketed values
  for the cases of  EoSs considered here.

  The necessary condition we  seek translates to
  $\tau_{\rm QN} > \tau_{\rm GW}$ which yields 
  $P_0  < 2.4~ {\rm hrs}\times  \tau_{\rm QN, 8}^{3/8} (M_{\rm NS, \odot}M_{\rm WD, \odot})^{1/3}/M_{\rm T, \odot}^{1/8}$ (for
  $\tau_{\rm QN}=10^8$ years). Thus   only 
 NS-WD systems born with periods $P_0$ in the  few hours range can be considered serious candidates for QNe-Ia in this scenario. 
 However,  evolution calculations of relevant IMXBs show that for progenitor mass range 
 $3.5 < M_{\rm don.}/M_{\odot} < 5$, the final binary
 periods are   days not hours (Tauris et al. 2000). 
It  might be the case  that  IMXBs which lead to  QNe-Ia are those who have experienced  additional angular momentum
 loss (i.e. orbit shrinkage). An interesting possibility is that 
  a small fraction of the transferred mass from the donor  form a circumbinary disc (CD). Evolution calculations in the context of  Cataclysmic Variables  (Spruit\&Taam 2001) and Black Hole IMXBs (Chen\&Li 2006) show that  a system with initial period of a few days reach final orbital periods of a few hours
    when a CD (which enhances the mass-transfer rate) is taken into account. In principle we  expect similar results for
    IMXBs described here if CDs are taken into account.
    Systems with CDs require additional angular momentum loss which is not considered in our model.
     Further studies on the evolution of such IMXBs are needed.

 While we do not expect QNe-Ia via S1 to be very common in today's universe (if these are related to $\eta_{\rm QN}$),
  it might have been different in the early universe  where a  boost in star formation rate  has been suggested.
    An order of magnitude estimate  of QN-Ia rate for this path  is
        $\eta_{\rm QN-Ia}  \sim
     \epsilon_{\rm IMXB} \times \eta_{\rm cc, 0}\times  \alpha_{\rm SFR}$
      where $\eta_{\rm cc, 0}$ is the current core-collapse rate 
       and $\alpha_{\rm SFR}$ is the boost in star formation rate at $1 < z < 2$.
       Here
       $\epsilon_{\rm IMXB} \sim  \epsilon_{\rm IMXB,CC}\times  \epsilon_{\rm IMXB,QNIa}$
where $\epsilon_{\rm IMXB,CC}$ is the fraction of CC which lead to an IMXB (of the order of $10^{-3}$; Pfahl et al. 2003),
while 
        $\epsilon_{\rm IMXB,QNIa}$ is the percentage of IMXBs  experiencing a QN-Ia.
        Assuming  current SD rate to be $\eta_{\rm SD, 0}\sim  \eta_{\rm cc, 0}/ 10$, we get
        $\eta_{\rm QN-Ia}  \sim  \epsilon_{\rm IMXB,QNIa}\times \eta_{\rm SD, 0}\times  \alpha_{\rm SFR, 100}$
         where  $\alpha_{\rm SFR, 100}$ is  in units of 100. The  QN-Ia rate would exceed
          the SD rate, $\eta_{\rm SD}$, if $\epsilon_{\rm IMXBs,QNIa} \ge 1\times  \eta_{\rm SD}/\eta_{\rm SD,0}$.
           For QN-Ias to be significant (compared to SD channel) at high redshift, we require $\epsilon_{\rm IMXB,QNIa} \sim 1.0$ which means that
virtually  all IMXBs should lead to  QN-Ias.

 In summary, unless the SD rate decreases drastically
 at higher redshift (i.e. $\eta_{\rm SD} << \eta_{\rm SD,0}$), it remains a challenge for QN-Ias   in this scenario (i.e. S1))
          to become statistically significant at any time.  Furthermore, CDs of the mass required in our model have yet to be confirmed observationally  (e.g. Muno\&Mauerhan 2006).
                 Another downside to this scenario is the $\tau_{\rm QN} >> \tau_{\rm GW}$ regime
          if most  NSs end up with a  magnetic field that is too small ($\le 10^8$ G). This case cannot be excluded and it would imply
            that the available time window for QN to occur when the system is still at $a=a_{\rm RL}$ is even shorter.

   \subsection{MNS-COWD systems with mildly recycled NS}    
    \label{sec:S2}

   The second scenario  (namely, S2) is that of  a tight binary system  with 
      the NS born massive (ideally  close to $M_{\rm NS,c}$) but not necessarily fully recycled (see \S 3.7 in van den Heuvel 2011 for massive NSs in binaries).
         In this case,  as long as the NS does not accrete as the system evolves towards
      $a_{\rm RL}$,  the QN explosion will most likely occur shortly after the WD overflows its RL, driving
       the NS mass (or core density) above $M_{\rm NS,c}$ (above the deconfinement value).  This means that $\tau_{\rm QN}\sim \tau_{\rm WD}$ in S2 
        which is a more universal and realistic result  since it is independent of the NS initial
        period and magnetic field.     
     
       It is well-known that a  NS+COWD system with a non-fully recycled NS
 can form in a very tight binary  via a CE (e.g. Ferdman et al. 2010).
  Since most  NS+CO tight binary systems form via the CE phase,
   this scenario provides better statistics for QN-Ia rate than the fully recycled NS one.
  Still, this would require that an important percentage of these
   lead to  MNS via accretion.  So far it appears that  MNS
      in CE channels  must have been born
     massive (e.g. Tauris et al. 2011) in which case their statistics would be tied to $\eta_{\rm QN}$ not to $\eta_{\rm CC}$.

\subsection{The Accretion-Induced-Collapse scenario}      
 \label{sec:S3}
 
  The possibility of merging of  CO WDs as SNe-Ia progenitors has been investigated (see Livio 2000 for a review).
 The outcome of these simulations is an inward propagating flame that converts the accreting CO WD  into an ONeMg WD. This star is gravitationally unstable and undergoes an AIC\footnote{Accreting CO-ONeMg systems where the ONeMg WDs may have formed directly from an $\sim 8$-10$M_{\odot}$ progenitor star are  also viable candidates for QNe-Ia.} to form a NS (e.g. Nomoto\&Iben 1985; Saio \& Nomoto 1985; Fryer et al. 1999).
  Modern simulations of  WD-WD  mergers suggest that  AIC as being the most likely outcome  (e.g. Saio \& Nomoto 2004; Yoon et al. 2007). 
If a Type I supernova is to follow from merging WDs, a thick disk must be formed as an intermediate stage in the merging process, with transfer from the disk onto the central degenerate dwarf occurring at a rate sufficiently less than Eddington that a deflagration induced by carbon burning occurs. Thus, the outcome of the merging of two massive CO degenerate dwarfs is not trivially a Type I supernova explosion. 
  Detailed 2-dimensional axisymmetric simulations of AIC  (Dessart et al. 2006\&2007) find that the AIC of white dwarf forms a $\sim 1.4$-1.5$M_{\odot}$
   NS, expelling a modest mass of a few  $10^{-3}M_{\odot}$  mostly through a neutrino-driven wind. A quasi-Keplerian accretion CO-rich disk with
    mass $\sim$ 0.1-0.5$M_{\odot}$ forms around the newly-formed proto-NS.

  The AIC of a WD to a NS releases significant binding energy.
The gravitational mass of the resulting NS is  the Chandrasekhar mass minus the gravitational binding energy of the NS.
Typically assumed numbers are that an accreting ONeMg WD, if pushed to AIC (at a mass of $1.44M_{\odot}$), leaves behind a $1.25 M_{\odot}$ NS. 
In this context,  the feasibility of a QN-Ia  relies heavily on: (i) the formation of  very massive $>1.3M_{\odot}$ ONeMg;  (ii) subsequent
             rapid accretion onto the $\sim 1.25M_{\odot}$ NS following AIC of the ONeMg WD; (iii)  A companion that should provide enough mass
             to first trigger the AIC then 
             to increase the NS mass from $\sim 1.25 M_{\odot}$ to $M_{\rm NS,c}$ and eventually evolve into a CO WD. 
             
             Let us consider a WD-WD system which consists of an ONeMg WD close to the $1.44M_{\odot}$ and a donor
              CO WD.  Since the AIC of the ONeMg WD  leads to an $\sim 1.25M_{\odot}$, the donor  companion
               must be massive enough  to provide enough mass to push the NS to $M_{\rm NS,c}$
               and leave behind a  CO WD. Taking into account the binding energy lost  during accretion from the companion,
              it is hard to imagine how accretion from a $\sim 0.6M_{\odot}$ WD companion can push
              the NS to  even the very low critical mass of $1.6M_{\odot}$, making the QN unlikely\footnote{See however Xu (2005) for alternative mechanisms for the formation of quarks stars via the AIC channel.}.  The need for a massive  donor WD  ($>> 0.6
M_\odot$) means that  the mass ratio is  sufficiently high that the two system will
merge.   If it turns out that such a merger leads to a MNS engulfed in
a degenerate CO-rich torus (e.g Yoon 2007), then  a QN-Ia is a possibility;  although this case violates the $q\le 0.5$ constrain of our model,  we consider it here for completeness.

            There is one more channel worth mentioning. 
            Belczynski \& Taam (2004) find that even if the ONeMg WDs were formed at a reasonably lower mass ($\sim 1.2 M_{\odot}$), some would  still be pushed over the Chandrasekhar mass limit\footnote{The formation of massive 
             ONeMg WDs might be challenging   for single star channel; the initial-final mass relationship
            derived by Meng et al. (2008) suggests that massive ONeMg WDs may only form for single stars at significantly super-solar metallicities (it means
            mostly in today's universe). }. They argue that  the last CE episode results in the formation of not only WD ($\sim$40\%) but also low-mass He star  ($\sim$40\%)  secondaries (see Table 1 in Belczynski \& Taam 2004). Either the WD or the helium star companion fills its Roche lobe and starts transferring material to the ONeMg WD. The AIC interrupts the mass transfer because of the loss of binding energy of the collapsing dwarf. However, in the case of a  helium star donor, mass transfer may restart on a short timescale, as nuclear expansion of the helium star is faster in bringing the system to contact than gravitational waves in the case of a WD donor. The helium star donors eventually lose sufficient mass to become low-mass  hybrid WDs (with a  carbon/oxygen in the core surrounded by a helium). 
            
            In principle, the helium star donor channel could lead to
 an AIC of a massive ONeMg WD while providing sufficient mass reservoir to form a massive enough NS to undergo a QN (see also Taam 2004
 and references therein).   The helium-donor case is the preferred scenario if the resulting low-mass hybrid WD could undergo a detonation
 following impact by the QNE (i.e. if  compression are high enough to achieve WD densities $> 2.5\times 10^7$ g cm$^{-3}$ and
 trigger burning under relativistic degenerate conditions).

  As a subset of AIC, the rate of QNe-Ia via this
      channel  depends on the AIC rate.  However, because AIC has never been observationally identified, its rate is uncertain.
  Theoretical estimates of the rate of AICs are also quite uncertain. 
  Based on r-process nucleosynthetic yields obtained from previous simulations of the
  AIC of white dwarfs, Fryer et al. (1999) inferred rates ranging from $\sim 10^{-5}$ to  $\sim 10^{-8}$  yr$^{-1}$ in a Milky-Way-sized galaxy. 
   This result depends upon a number of assumptions  and the true rate of AICs could be much lower or much higher  than this value. 
    If  higher numbers can be confirmed,
   then  a small percentage of AICs leading to  MNSs  (with the subsequent QNe explosions under conditions described above)
   could make QNe-Ia statistically viable.

 A population that is  close to the NS-WD systems described here  is the 
   ``Ultra-Compact X-ray binaries"  population that contain
   CO WDs (e.g. Nelemans et al. 2010).  These have most likely evolved via the CE with the WD 
   probably sitting at $a_{\rm RL}$.  In particular if it can be shown that
   some UCXBs contain MNS then these would be potential QNe-Ia candidates.
   These we will investigated  elsewhere.

\subsection{Summary}

 To summarize this section,  we have presented three possibles progenitors
 of QNe-Ia.   S1 and S2 appeal to IMXBs that lead to a  tight MNS-COWD binary.
   S1 and S2 are related to massive star formation (rate) while S3 is related to  slightly lower  mass 
             star formation (since the NS forms from the AIC of a WD).
 We find the S1 scenario (fully recycled NS) the least likely progenitor  not only because
  of its extremely rare occurrence but also because of 
  the constraint it imposes on the MNS-COWD birth period ($P_0$ much less than a few hours).
   S2 should be considered a serious option if the Pop.II IMF favoured
  more massive stars.  
 There is  also the intriguing possibility of the AIC channel (S3) 
 which  might be the most  viable  statistically  if AIC rate is indeed very high;  which remains to be confirmed.
 S2 and S3 we argue  lead to relatively prompt
   explosions  (with delay time $t_{\rm delay}$ not exceeding a few Myrs; see \S \ref{sec:sfr})
   while S1 would lead to a longer  time delay.
  We mention that if it happens that mergers occur at smaller values of $q$ than considered here (i.e. $q < 0.5$),
   then S1 and S2 would be less likely since these depend heavily on stable mass-transfer; only S3
    would remain as a viable QN-Ia progenitor.
          Much uncertainty still remains regarding the formation and evolution of close binary stars
           in particular those evolving through a CE phase and/or a WD-WD path.  However, 
           if any of these scenarios could  lead to tight (i.e. with $a\sim a_{\rm RL}$) MNS-COWD systems  with NS  masses close to $M_{\rm NS,c}$
            then we might  have at hand a viable channel for QNe-Ia.

Despite these uncertainties,  we mention that binary evolutionary paths
 exist that could lead to   compact binary systems with MNS and a CO WD  as described above.
 For example, the  evolutionary path C  shown in Figure 1 in Stairs (2004) 
  lead to a binary system closely resembling  PSR J1141$-$6545 with a NS mass
  of 1.3$M_{\odot}$ and a WD mass of 1$M_{\odot}$ (see Table 1 in Stairs 2004).
   This system has a birth period of 0.2 days which is within the conditions described for the S1 scenario
   above.     Another candidate that could potentially evolve to an S1 case is  PSR J1802$-$2124 which consists 
    of an $\sim 1.24M_{\odot}$ mildly recycled ($\sim 12.6$ ms) NS  and a $\sim 0.78M_{\odot}$
     WD (Ferdman et al. 2010). Accretion onto the NS   could in principle  increase  its mass  to $M_{\rm NS,c.}$ (i.e.  recycling it)
      and trigger a QN event and subsequently a QN-Ia.

\section{Discussion}
  
   \subsection{Other sub-Chandrasekhar models}

   As discussed in the Introduction,  sub-Chandrasekhar models 
   can be classified as: (i) edge-lit (with helium layer) single-degenerate sub-Chandrasekhar mass
   explosions (Kasen\&Woosley 2011); core-lit (w/o helium layer) single-degenerate sub-Chandrasekhar mass
   explosions (Sim et al. 2010); (iii) core-lit  sub-Chandrasekhar mass remnants from mergers of roughly equal-mass CO WDs (van Kerkwijk et al. 2010).
   These involve $>0.9M_{\odot}$ WDs and are all powered purely by $^{56}$Ni decay.
    The lower mass WDs in theses model   means a deflagration might not necessary to produce the observed IMEs. 
    Some are more successful in reproducing observed SNe-Ia than others.

 There are  fundamental differences between our model and those described above.
 E.g.:   (i) The compression and heat deposition induced by the impact of the QNE puts the WD  in a regime 
``mimicking" massive WDs ($\rho_{\rm SWD}> 10^8$ g cm$^{-3}$) despite the much lower mass WD involved ($M_{\rm WD} < 0.9M_{\odot}$). 
Our model, involves truly  sub-Chandrasekhar mass WDs  at explosion.
 As we argued in \S \ref{sec:compression},  the shock from the QNE impact could  in principle reach deep 
 into the WD core to trigger an inside-out (i.e. a centrally ignited) explosion; 
 (ii)  In our case no helium layer is necessary. I.e.  the explosion is independent of accretion onto the WD.
 Nevertheless,  as discussed in \S \ref{sec:S3},
   one of the AIC channels would lead to a MNS surrounded by a hybrid HeCO WD. If these system experience a QN explosion
   of the MNS then they would have some distinct (photometric and spectroscopic)
   properties given the extremely low-mass WD ($<0.3M_{\odot}$) and the presence of helium;
  (iii) The additional energy source (i.e. the spin-down power) would affect the evolution of the fireball and
   the resulting light-curve.

\subsection{The connection to star formation ?}
\label{sec:sfr}

Type Ia supernovae are seen to occur in early type (elliptical) galaxies and in younger stellar populations. 
Observations have shown that they are more prevalent in star-forming late-type galaxies than in early-type galaxies (Oemler \& Tinsley 1979). 
Young galaxies host roughly two times more SNe-Ia than early galaxies (Nomoto et al. 2000) because SNe-Ia are slightly more efficiently produced in younger stellar populations (Bartunov et al. 1994).
The mean luminosities of SNe-Ia observed in spiral galaxies are clearly higher than those of elliptical galaxies (Wang et al. 2008). 
  A very significant factor here is the absence of the brightest SNe-Ia in elliptical and S0 galaxies. 
 The current explanation for these observations is that there are prompt (delayed by  $\sim 200$ to $\sim 500$ Myr from the onset of star formation; Oemler \& Tinsley 1979; Raskin et al. 2009) and delayed (tardy; $>$ 1 Gyr) SN-Ia explosions (e.g. Ruiter et al. 2009 and references therein). The prompt component is dependent on the rate of recent star formation, and the delayed component is dependent on the total number of low-mass stars. The combination of these two components is believed to form the overall observed SN-Ia rate (Hamuy et al. 2000; Sullivan et al. 2006; Wang et al. 2008; Sullivan et al. 2010).

In S2 and S3, QNe-Ia  might occur shortly 
 after star formation with a delay   associated  with the  donor's main-sequence lifetime.
 Specifically, $t_{\rm delay}\simeq \tau_{\rm M_2}+\tau_{\rm GW} \simeq    \tau_{\rm M_2} 
   \sim 3\times 10^{8}\ {\rm yrs}\ (4 M_{\odot}/M_{\rm 2})^{2.5}$ 
 which gives\footnote{Mass exchange during the binary evolution makes it hard to pin-point the exact 
 range of WD progenitor mass. However, in our model this exchange may be minimal
  given the mass of the primary   ($20M_{\odot} < M_{\rm prim.} < 40M_{\odot}$; for the S2 channel) required to form a massive
   NS at birth. The primary would explode as a SN  very shortly
   after the binary's formation. In this case we do not expect much interaction and mass exchange
    with the secondary until it evolves to the red giant phase.   For a $\sim 0.6M_{\odot}$ WD this justifies the 
      secondary's mass   range we adopt of  $4M_{\odot} < M_{\rm 2} < 7M_{\odot}$ (e.g. Tauris 2011).}  $70\ {\rm Myr} < t_{\rm M_{\rm delay}} < 300$ Myr; 
 this assumes  $\tau_{\rm QN}\sim \tau_{\rm GW} < \tau_{\rm M_2}$.

 A burst in massive star formation at high redshift combined with  a slightly heavier IMF of Pop. II stars
  would increase the formation rate of MNSs and also probably of massive CO WDs.
   The increase in massive CO WDs could lead to an increase in  ONeMg WDs numbers via accretion
  processes described in \S \ref{sec:S3} (at high-redshift and low-metallicity,  direct formation of ONeMg WDs from single stars is heavily reduced;  Meng et al. 2008).  
  The suggested peak in star formation rate   at redshifts $1 < z < 2$ (e.g. Hughes et al. 1998;
   Madau et al. 1998; Pettini et al. 1998; Dickinson et al. 2003) combined with a
    heavier IMF of Pop. II stars would  make S2 and S3 channels highly plausible and may be prominent  in the early universe.  
   This means that the QN-Ia rate could  peak at  $0.75 < z < 1.75$
    if they occur on average $<$ 300 Myr after the onset of star formation  (Wright 2006). 
       Although the rate estimates given in this work are still subject to substantial uncertainties.

 \subsection{Plausible Implication to Cosmology}
\label{sec:cosmology}

SNe-Ia have been successfully used as standardizable (Phillips 1993) distance candles and have provided the first indication for an accelerating Universe and the need for Dark energy (Riess et al., 1998; Perlmutter et al., 1999). Effectively, 
they provided evidence for a universe that experiences an accelerated expansion since the time when it was about half of its present age. At that time, the predicted dark energy took over the kinematics of our universe that was ruled by the matter contribution before. 
 This conclusion was based on a sample of 
over a hundred of  nearby SNe-Ia that have been studied, revealing considerable homogeneity.
However   some fascinating differences between SNe-Ia do exist (e.g.
Phillips 2012) and in particular 
for those interested in using SNe-Ia as cosmological distance indicators, the most troubling of the peculiar objects are the 2006bt-like SNe.

 SN 2006bt was observed to have a fairly broad, slowly decaying light curve, indicative of a luminous supernova.
However, it displayed intrinsically-red colors and optical spectroscopic properties that were more like those of fast-declining, low-luminosity events
(it was also lacking the $i$-band shoulder).  Although SN 2006bt appears to have a somewhat odd light curve, it is still a relatively good standardizable candle.
The intrinsically-red color evolution of the SN caused standard light-curve fitting programs to significantly overestimate the dust reddening\footnote{Light-curve fitters must correct for the fact that redder supernovae are dimmer. This is due to a combination of an intrinsic color-luminosity relation (faint supernovae are intrinsically red; Riess, Press, \& Kirshner 1996), and reddening due to dust.}. 
This, despite the fact that SN2006bt occurred in the outskirts of a galaxy, showing no sign of dust absorption.
 All light-curve fitters correct for its red color by effectively brightening its apparent magnitudes.
    This brightening correction followed by standard  calibrations techniques could either over-estimate or
    under-estimate the true magnitude. Foley et al. (2010)  developed a Monte Carlo simulation to assess the impact of contamination of a population of SN 2006bt-like objects in a SN-Ia cosmological sample. Using basic simulations, they showed that SN 2006bt-like objects can have a large impact on derived cosmological parameters.
 It  can be seen from their Figure 12, that  10\%  contamination of SN 2006bt-like objects in the nearby ($z < 0.1$) and full sample
 increases the scatter of a SN-Ia Hubble diagram and systematically bias measurements of cosmological parameters (see also their Table 3).
 
  We have already noted the intriguing similarities between SN2006bt-like (and other peculiar SNe-Ia) objects and QNe-Ia (see \S \ref{sec:candidates}). 
In particular we noted and showed that some QNe-Ia light-curves could be mistaken for  standard SNe-Ia of higher
$^{56}$Ni content (see Fig. \ref{fig:fig2}).  Since photometry and spectroscopy are not necessarily
linked in QNe-Ia, light-curve fitters  would be confused by these.
 They would  apply  brightening corrections and  standard  calibrations techniques that could either over-estimate or
    under-estimate the true magnitude of these QNe-Ia.
   Once the QN-Ia light-curve is derived/computed in detail, 
 a study similar to Foley et al.(2010) should
 be performed in order to assess plausible QNe-Ia contamination and the implication to  SN-Ia Hubble diagram. 
   For now, if the analysis of Foley et al. (2010) were any indication, we are
 tempted to speculate that QNe-Ia if they exist (in particular at high redshift)
  might  systematically bias measurements of cosmological parameters.

Are QNe-Ia contaminating the high-redshift  SNe sample?
Observations show that    the  most luminous SNe-Ia (those with the broadest lightcurves),  favor star forming hosts and
   occur only in late-type galaxies (Hamuy et al. 1996). Since star formation increases by a factor of 10 up to redshift 2 (e.g. Hughes et al. 1998;
   Madau et al. 1998; Pettini et al. 1998; Dickinson et al. 2003), it is expected that the mix of supernovae will change with redshift. 
   Howell et al. (2007) finds the fraction of  broad light-curve SNe increases with redshift and seems
to agree with the idea that 
as star formation increases with redshift, the broader light-curve SNe-Ia associated with a young stellar population
 make up an increasingly larger fraction of SNe-Ia.
   We have already argued that spin-down powered QNe-Ia should be associated with bright and broad
   light-curves  and should be linked to star forming regions (the  delay time
  between formation and explosion should not exceed a few Myrs for S2 and S3; see \S \ref{sec:sfr}) which means they
  should exist at high $z$.  
  In particular, QNe-Ia with $\zeta_{\rm sd}$ closer
   to 1 would be extremely luminous and should be easily detected (if not dominant) at high redshift. 
   In addition, as mentioned in \S \ref{sec:sfr}, a burst in massive star formation at high redshift combined with  a slightly heavier IMF of Pop. II stars
  would increase the formation rate of MNSs and also probably that of massive CO WDs.   
  The increase in the mass of CO WDs  could, overall, make QNe-Ia  at high-redshift produce more $^{56}$Ni 
  which should increase their brightness.

   If observed  luminous high-redshift SNe (or at least a percentage of them) are truly luminous QNe-Ia (i.e.  powered by   spin-down), then 
      these should be removed from the sample before calibrations\footnote{If these were powered purely by $^{56}$Ni decay then this mixing 
    is not necessarily problematic for cosmology since light-curve shape, color corrections 
    and correction for host galaxy properties allow all supernovae to be corrected to the same absolute magnitude;
    thus avoiding  systematic residuals with respect to the Hubble diagram.} are made.
    Unfortunately, at such high-redshift  (in fact for any SN at $z>0.3$) the $i$-band is redshifted out of the optical, thus 
   making the identification as QNe-Ia very challenging.
   Nevertheless, one could in principe rely on unique features of QNe-Ia light-curves and spectra to differentiate
   between $^{56}$Ni powered SNe and QNe-Ia at high-redshift (see \S \ref{sec:predictions}). 
    If not, then if  QNe-Ia were to account for a  percentage (may be as low as $\sim$ 10\%) of the 
   SN-Ia sample at  high-redshift (in our rough estimates at $0.5 < z < 1.75$),   one wonders if  these
   could systematically bias measurements of cosmological parameters as to allow 
   for other cosmologies; and maybe explain away the need for  Dark energy? 
  This is of course a daring and highly speculative conclusion since it ignores the constraints from 
    the ``concordance model"\footnote{The values of $\Omega_{\Lambda}$ and $\Omega_{\rm M}$ are  confirmed also from the examination of the cosmic microwave background (CMB) and galaxy clusters. The consistence of these three methods is known as Cosmic Concordance. The position of the first Doppler peak as well as the comparison of the peak amplitudes for different multipole moments in the CMB angular power spectrum indicate a flat Universe and constraint the sum of $\Omega_{\rm M}$ and $\Omega_{\Lambda}$ (e.g. Spergel et al. 2003
and references therein). $\Omega_{\rm M}$  is constrained by the evaluation of the mass of galaxy clusters (e.g. Carlberg et al. 1998).} (see however Kroupa et al. 2010).

\section{Predictions and conclusion}
\label{sec:predictions}

We proposed a new model for Type Ia SNe involving a QN going off
in a tight MNS-COWD binary.  The impact of the dense relativistic  QNE on the
WD could lead to compression and heating which, under appropriate conditions,
 should lead to the ignition and detonation of the WD; the QN-Ia. A particularity
 of our model is the spin-down power from the QS (the QN compact remnant) 
 which provides an additional energy input (besides the $^{56}$Ni decay) to make the QN-Ia fairly
 bright.  Our preliminary calculations of QNe-Ia light-curves suggest that 
 these should somewhat deviate from the Phillips relationship although some
 are close enough to standard (i.e. $^{56}$Ni-powered) SNe-Ia that they could easily be confused as such.
 Another particularity of our model, is that the photometric and spectroscopic properties
 are not necessarily linked. Light-curve fitters ``stumbling on"  a QN-Ia would find
 a discrepancy between the light-curve and the spectrum at   peak and would try
 to correct for it by incorrectly brightening or dimming the object. This we argue
 could systematically bias measurements of cosmological parameters if QNe-Ia at high-redshift are numerous and bright enough to be included
 in the cosmological sample.

   Some features/predictions of our model are:
       
       \begin{enumerate}[(i)]
       
       \item   We start by pointing out that we expect the QN proper 
       to be  much less luminous (in the optical) than the QN-Ia.
                Most of the QN energy  is in the Kinetic energy of the QNE so unless the QNE interacts strongly with its surroundings it will not be optically bright; the QN  will be dwarfed  (in the optical) by  the QN-Ia  given the low-density  expected in the binaries considered here. 
         In much denser environments,  the collision between the QNE and the surroundings lead to 
         extremely  bright  QNe (Leahy\&Ouyed 2008;  Ouyed\&Leahy 2012).

     \item   Given the QNE kinetic energy ($\sim 10^{52}$ ergs),  QNe-Ia
        should be associated with cavities  (i.e. they would carve out bubbles) much larger than those expected from 
         Type IIs and standard Type Ias.
         
           \item      The expanding neutron-rich QN ejecta would have processed mostly heavy elements  with 
 atomic mass $A > 130$ (Jaikumar et al. 2007). The $\sim 10^{-3}M_{\odot}$ in the QN ejecta 
 provides substantial amount of $A>130$   to contaminate 
   their environment  (and thus QNe-Ia environments) with such elements.

        \item     The nature of the GW signal from a QN  has been computed in Staff et al. (2012).  Prior to the QN explosion proper, the NS-WD objects  described here would  also be a source of detectable  signals
  since we expect them to  be more common than NS-NS and/or NS-BH systems. 
   Expectedly, GW signals from SD and/or DD channels would be distinct from those
  from QNe-Ia. In the QN-Ia model, we expect a delay between the  GWs signalling the QN proper 
   and the GWs signalling the explosion of the WD.  The delay is of the order
   of a fraction of a second and is a combination of the time it takes the QNE to reach
   the WD and the burning time of the WD.

           \item  Unlike other models of SNe-Ia so far proposed in the literature, the QN leaves behind a compact star.
   The compact remnant, in our model, would be a radio-quiet 
  quark star (an aligned rotator; Ouyed et al. 2004 and Ouyed et al. 2006)  
   with specific X-ray signatures (Ouyed et al. 2007a\&b). 
   
   \item   The spin-down luminosity of the resulting quark star  (Staff et al. 2008)
    would result in the formation of a wind nebulae (much like a pulsar-wind nebula) which is another unique
    feature of our model.  Association of a SN-Ia with a pulsar-wind nebula would
      strongly support our model.  In particular, given the similarities between QNe-Ia and 
      and peculiar SNe-Ia (see \S \ref{sec:candidates}), it would be interesting to search for signatures of a pulsar-wind nebula (or even extremely large cavities) in peculiar SNe-Ia. One could search   for possible signatures of  any  asymmetries 
    in the propagating ejecta e.g.,  by using polarization measurements taken at early times (prior to maximum light; e.g. Wang\&Wheeler  2008).

\item If the high-redshift SNe-Ia are truly spin-down powered QNe-Ia, these would  lack (or show a weak)
second maximum in the $i$-band.  Although the $i$-band would be shifted from the optical
one could in principal perform this exercise in the infra-red (which should be within the 
 JWST reach at $z\sim 1$).
 
 \item   Applying  Arnett's law to a QN-Ia, as we have said, would lead to an overestimate of the
 true $^{56}$Ni yield. In QNe-Ia, the $^{56}$Co yield one would infer from the later times
 ($>> 8.8$ days)  would be much smaller than those obtained around peak.
 
\item Finally, we suggest a few QNe-Ia candidates among historical Galactic Type-Ia remnants.
 According to predictions in the SD models, the companion star (i.e. the donor star) should survive the explosion and thus should be visible in the center of Type Ia remnants.   A direct detection of a surviving donor star in a Galactic Type Ia remnant would substantiate the SD channel for at least one system.   Among the known Galactic remnants (e.g. Tycho Brahe's SN,
  Kepler's SN, SN 1006) none shows undeniable presence of  a surviving companion.  For example,
  the well studied  SN 1006 seems to be lacking a surviving donor star  (Kerzendorf et al. 2012; Gonz{\'a}lez Hern{\'a}ndez et al. 2012).   Recently, Schaefer \& Pagnotta (2012) reported  that the central region of the supernova remnant SNR 0509$-$67.5 (in the Large Magellanic Cloud) lacks an ex-companion down to very faint magnitudes.  
   While  the DD scenario might be an alternative  progenitor\footnote{If one assumes that the observed type Ia SNe
are a combination of DD and SD events, it would be quite a coincidence
that most of the nearby, well-studied, SNe-Ia had  DD progenitors.},
  we argue that for those remnants where the SD can be ruled out  through  deep imaging observations, i.e. those with 
clear  lack of any ex-companion star, the QN-Ia avenue should also be explored.  

   SN 1006 and SNR 0509$-$67.5 are particularly interesting and should be considered potential QN-Ia contenders. 
   If  future deep monitoring of these systems  reveal a radio-quiet\footnote{The compact remnant (i.e. the QS) in the QN model is born as an aligned rotator (Ouyed et al. 2004; Ouyed et al. 2006). Thus explaining why a radio pulsar has not been detected in this remnant.} compact remnant (the QS) emitting in the X-rays then one can make a strong case for a QN-Ia. Another clue to look for is 
   a cavity carved out by the QN ejecta prior to the QN-Ia; in our model, the QN explosion proper (which explodes first)
    creates the cavity into which the WD explodes following impact from the QNE.    
      In this context, we should mention 
  another historical remnant of interest to our model namely, RCW86. A  self-consistent explanation of   the Infra-red, X-ray, and
optical observations in this object, presumably  requires  an explosion into a cavity created
by the progenitor system (Williams et al. 2011).     This hints at the SD channel
   where  the progenitor might have carved a wind-blown bubble. 
   However, if future measurements can rule out a main-sequence or a giant companion in RCW86, 
   then this would lend support for the QN-Ia assuming that  wind-blown bubbles cannot be formed in 
   DD models or in SD models in general.

     \end{enumerate}

  To compare to detailed observations, it is necessary to perform detailed
multi-dimensional, hydrodynamical simulations of the relativistic QNE impacting the dense
degenerate WD  under
the settings described in this work, and couple
the results with a nuclear network code to properly capture the relevant nucleosynthesis
 during WD burning and subsequent ablation.  In particular these simulations 
  would be important in assessing how much of the QN shock 
  would pass through the WD and how much will go around it.    Furthermore, a detailed study
   of how the spin-down energy is deposited and dissipated in the WD material remains to be done
    and its implications on the morphology of the light-curve to be shown.
     Finally, preliminary 1D simulations  of the QN (Niebergal et al. 2010) indicate that our working hypothesis -- the QN as a detonative
 transition from neutron to quark matter inside a NS -- might be valid.
  If this is borne out by more sophisticated simulations, we will have found potentially a new engine for 
   luminous (spin-down powered) sub-Chandrasekhar  mass ($<0.9M_{\odot}$) SNe-Ia plausibly hidden among observed SNe-Ia  at   high redshift.
   This highly speculative, but 
    exciting, possibility  should make our model for QNe-Ia (and the QN proper) worth pursuing.


\begin{acknowledgements}
We thank P. Jaikumar, D. Welch, T. Tauris, J. Frank and B. Zhang for helpful discussions.  
We thank D. Leahy for comments on an earlier, letter version of this manuscript. 
The research of R. O. is supported by an operating grant from the
National Science and Engineering Research Council of Canada (NSERC). 
This work has been supported, in part, by grant  NNX10AC72G from NASA's ATP program.
\end{acknowledgements}

\appendix

\section{Relativistic Shock Jump Conditions}

A shock is described by three jump conditions that express the continuity of mass, energy, and momentum flux densities, respectively, in the shock frame (Landau \& Lifshitz 1959). When the QNE encounters the  WD (i.e. a density jump $n_{\rm QNE}/\rho_{\rm WD}$),
 a reverse shock (RS) is driven into the cold QNE, while a forward shock (FW) propagates into
 the cold higher density WD material.    Therefore, there are four regions separated
by the two shocks in this system: (1) unshocked cold WD matter, (2) forward-shocked hot WD matter, (3) reverse-shocked QNE,
and (4) unshocked cold QNE.

We denote $n_i$, $e_i$ and $p_i$ as the baryon
number density, energy density and pressure of region ``$i$" in its own rest
frame respectively; $\Gamma_i$ and $\beta_i$ are the Lorentz factor and
dimensionless velocity of region ``$i$" measured in the local medium's rest frame
respectively.
The reverse- and forward-shock jump
conditions simply state the conservation of energy, momentum, and
particle number across the shock, which is equivalent to the
continuity of their corresponding fluxes. We assume the equations of state for regions 2 and 3   to be
relativistically hot and regions 1 and 4 to be cold. In regions 2 and 3 then $\rho_2c^2 \ll
p_2$ and $\rho_3c^2 \ll p_3$ and the adiabatic index  is $4/3$, implying $p_i = e_i/3 = w_i/4$; $w$ being the enthalpy.
In region 1 we have $\rho_1 = w_1/c^2$, $p_1 = e_1 = 0$ and $\Gamma_1 = 1$
while region 4 describes the QNE.  This leaves eight unknown quantities: $\Gamma$, $n$ and
$e$ in regions 2 and 3, as well as the Lorentz factors of the
reverse shock, $\Gamma_{\rm RS}$, and of the forward shock, $\Gamma_{\rm FS}$.
Correspondingly, there are eight constraints: three from the shock
jump conditions at each of the two shocks, and two  at the contact
discontinuity (pressure equilibrium and velocity equality
along the contact discontinuity) : $e_2 = e_3$ and $\Gamma_2 = \Gamma_3$. 
The equations describing the jump
conditions for the forward shock becomes (Blandford \& McKee 1976)
\begin{equation}
\frac{e_2}{n_2m_pc^2}=\Gamma_2-1,  \,\,\,\,\,\,
\frac{n_2}{n_1}=4\Gamma_2+3,
\end{equation}
where $m_p$ is the proton mass. The relevant equations for the RS can similarly be derived. 

Under the conditions specified above, the solution of the jump equations depends
only on two parameters (e.g. Sari\&Piran 1995): $\Gamma_4=\Gamma_{\rm QNE}$ and $f = n_4/n_1=n_{\rm QNE}/n_{\rm WD}$
(in our case $n_4=n_{\rm QNE}$ is the number density of the QNE and $n_1=n_{\rm WD}$ is the
number density of the WD). The number density in the shocked WD (SWD) material/region is then $n_{\rm SWD}=n_2$
with $n_{\rm SWD}/n_{\rm WD}$ given in equation above.
 In the co-moving frame of the "shocked" WD material (i.e. region 2)  $n_{\rm SWD}^{\prime} = \Gamma_2 n_{\rm SWD}$.

 The  Lorentz factor of the shocked WD material (i.e. region 2), $\Gamma_2$   can be shown to be 
  \begin{equation}
 \Gamma_2 = 
\left\{
 \begin{array}{rl}
\Gamma_4 & \mbox{if $n_4/n_1 < \Gamma_4^2 $ \ ,}
\\
 (n_4/n_1)^{1/4}\Gamma_4^{1/2}/\sqrt{2}  & \mbox{if $n_4/n_1 > \Gamma_4^2 $\ .}
\end{array}
\right.
\end{equation}   
 When the RS is Newtonian ($n_4/n_1 < \Gamma_4^2 $) it  converts only a very small fraction of the kinetic energy into thermal energy; in this case the Lorentz
  factor of region 2 (the shocked WD material) relative to the rest frame of the WD (i.e. region 1; also an external observer) is  $\Gamma_2\simeq \Gamma_{\rm QNE}$.   The relativistic RS  limit  ($n_4/n_1 > \Gamma_4^2 $)
  is the regime where most of the kinetic energy of the QNE is converted to thermal energy by the shocks (in this
  case $\Gamma_2\simeq (n_{\rm QNE}/n_{\rm WD})^{1/4}\Gamma_{\rm QNE}^{1/2}/\sqrt{2}$).
 
\label{lastpage}
\end{document}